\documentclass[10pt,a4paper]{article}

\usepackage[T1]{fontenc}
\usepackage[utf8]{inputenc}
\usepackage{amssymb,amsmath,amsthm,amsfonts}    
\usepackage{bm}                                 
\usepackage{mathtools}                          
\usepackage{stmaryrd}                           
\usepackage{breqn}                              
\usepackage[textfont=footnotesize,labelfont={footnotesize,bf}]{caption}
\usepackage{subcaption}
\usepackage{booktabs}
\usepackage[inline]{enumitem}
\usepackage{graphicx}
\usepackage{graphbox}
\usepackage[dvipsnames]{xcolor}
\usepackage[top=30mm,right=28mm,left=28mm,bottom=30mm]{geometry}
\usepackage[colorlinks=true,linkcolor=blue,urlcolor=blue]{hyperref}
\usepackage[affil-it,auth-sc]{authblk}
\usepackage[colorinlistoftodos,textwidth=25mm,textsize=footnotesize]{todonotes}
\usepackage{lipsum}                             
\usepackage[section]{placeins}
\usepackage{tikz}
\usetikzlibrary{positioning}
\usepackage{scalefnt}
\usepackage[
	backend=biber,
	style=numeric,
	citestyle=numeric-comp,
	maxbibnames=99,
	minbibnames=99,
	maxcitenames=2,
	mincitenames=1,
	giveninits=true,
	sorting=none,
	sortlocale=auto,
	natbib=true,
	url=false,
	isbn=false,
	doi=true,
	eprint=true
]{biblatex}
\newcommand{\nderiv}[3]{\frac{\partial^{\,#3} #1}{\partial #2^{\,#3}}}

\newcommand{\bzero}{\bm 0}
\newcommand{\ba}{\bm a}

\newcommand{\be}{\bm e}

\newcommand{\bg}{\bm g}

\newcommand{\bk}{\bm k}

\newcommand{\bn}{\bm n}

\newcommand{\bq}{\bm q}

\newcommand{\bu}{\bm u}

\newcommand{\bx}{\bm x}

\newcommand{\bA}{\bm A}

\newcommand{\bK}{\bm K}

\newcommand{\mC}{\mathcal C}

\graphicspath{{./}{./figures/}}                 

\title{Dynamics of elastic lattices with sliding constraints}

\author[1]{L. Cabras}
\author[2]{D. Bigoni\footnote{Corresponding author: e-mail: \href{mailto:name@unitn.it}{bigoni@unitn.it}; phone: +39\,0461\,282507.}}
\author[2]{A. Piccolroaz}

\affil[1]{Department of Industrial and Mechanical Engineering, University of Brescia, Italy}
\affil[2]{Department of Civil, Environmental, and Mechanical Engineering, University of Trento, Italy}

\date{}

\begin{document}

\maketitle

\begin{abstract}
This study investigates the impact of sliders -- constraints acting on elastic rods allowing for a transverse displacement jump while maintaining axial and rotational displacement continuity -- on the dynamics of a periodic elastic grid, including the effects of axial preload. The grid is linearly elastic and subject to in-plane incremental deformation, involving normal and shear forces and bending moment. The periodicity of the infinite grid permits a Floquet-Bloch wave analysis and a rigorous dynamic homogenization, leading to an equivalent prestressed elastic solid. The investigation is complemented by {\it ad hoc} developed F.E. simulations and perturbations with a pulsating Green's function. 
Results show that the sliders create band gaps, flat bands, and Dirac cones in the dispersion diagrams and generate macro-instability even for tensile prestress. The latter corresponds to the loss of ellipticity at the parabolic boundary in the equivalent elastic solid and provides a rare example of an almost unexplored form of material instability. Therefore, our results offer design strategies for metamaterials and architected materials showing reversible material instabilities and filtering properties for mechanical signals. 
\end{abstract}

\paragraph{Keywords}
Metamaterials \textperiodcentered\ 
Floqet-Bloch analysis \textperiodcentered\
Shear bands \textperiodcentered\
Ellipticity loss \textperiodcentered\
Material instabilities

\section{Introduction}
\label{sec:introduction}
Slider constraints implement transverse compliance (at the same time, maintaining continuity of axial displacement and rotation) in Euler-Bernoulli or Rayleigh rods, which are otherwise unshearable. Breaking unshearability leads to strong effects on structural mechanics, for instance, introduces tensile buckling in slender rods \cite{zaccaria_2011}. Architected materials, designed as a grid of axially preloaded elastic rods with sliders, were homogenized to obtain the response of an equivalent prestressed material \cite{bordiga_tensile_2022}. The equivalent material was shown to possess a completely bounded stability domain, while unboundedness for tensile prestresses always occurs when sliders are absent \cite{bordiga_2021}. Therefore, the use of sliders yields an elastic metamaterial which \lq fails' (i.e. develops shear bands, instead of breaking bonds as in \cite{nieves_analysis_2016}) for every load path, while \lq integrity' may be recovered by simply unloading. 

\begin{figure}[htb!]
    \centering
    \includegraphics[height=0.50\linewidth]{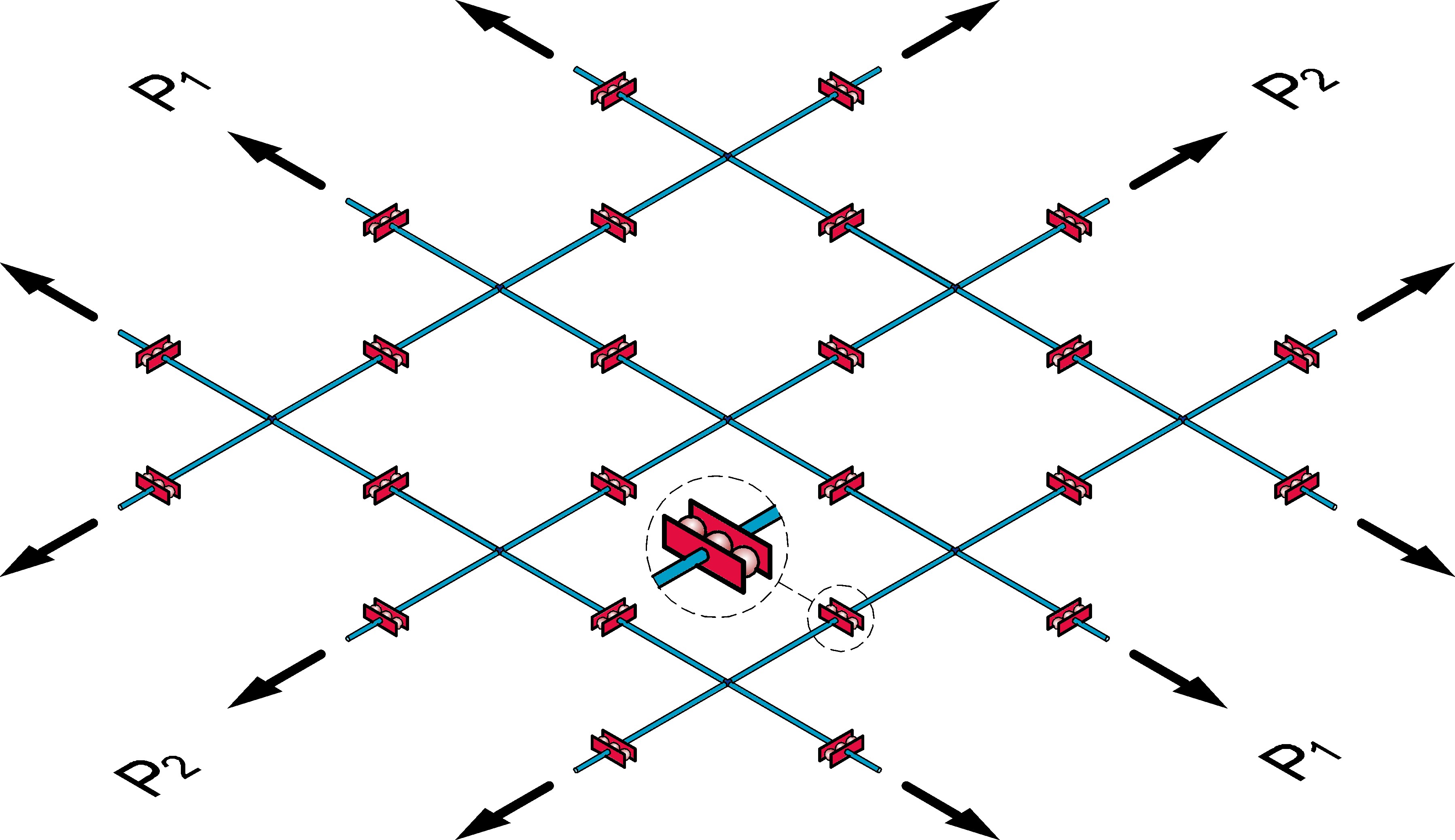}
    \caption{
    \label{slideroni} A grid of elastic and axially-preloaded rods, jointed through sliding constraints, provides an example of a metamaterial with a bounded stability domain (thus buckling even in tension). The fact that tensile preload may be applied precludes out-of-plane instabilities, so that the material can be used for structural membranes. In dynamics, the grid shows multiple band gaps, flat bands, and Dirac cones, while its equivalent homogeneous material provides an uncommon example of a material losing ellipticity at the parabolic boundary, thus evidencing \lq stress channelling', corresponding to shear bands parallel (orthogonal) to the direction of the tensile (compressive) applied preload, $P_1$ and $P_2$. 
    }
\end{figure}

Although many results have been obtained, the investigation into the mechanical effects of sliders is far from concluded, so that important effects may still have to be discovered. The present article contributes in this direction by analyzing the dynamics of a periodic grid of axially-preloaded elastic rods, each jointed at mid-span with a slider, Fig.~\ref{slideroni}. 
%

Dispersion surfaces are obtained through a Bloch-Floquet wave analysis, revealing band gaps, flat bands, and Dirac cones. Here, the sliders lead to the emergence of twin band gaps and localize the propagation in a direction parallel to the sliding direction when the stiffness of the sliders tends to vanish, so that a \lq floppy mode' is introduced in the kinematics. The analysis is complemented with F.E. simulations (of a portion of grid {\it ad hoc} implemented with a PML boundary), to be contrasted with results obtained using a Green's function perturbation technique (see \cite{bigoni_2012}), applied to an infinite body made up of the prestressed elastic material equivalent (through rigorous low-frequency, long-wavelength dynamic homogenization \cite{triantafyllidis_1993, willis_2009,  santisidavila_2016,  willis_2011}) to the elastic grid. 
When macroscopic bifurcation occurs, corresponding to loss of ellipticity in the equivalent material \cite{pontecastaneda_1989, avazmohammadi_2016, furer_2018}, shear band formation is observed in both the lattice structure and the equivalent material. The analyses reveal that the architected material designed in the present article loses ellipticity at the elliptic/parabolic boundary through the formation of shear bands aligned parallel (orthogonal) to the direction of the applied tension (compression). Materials behaving in this way were discovered by Pipkin \cite{pipkin}, who showed that they display \lq stress channelling'. They are useful to localize signals, but are extremely rare (one example being that illustrated in \cite{muretti}), so that a new design strategy towards these materials is introduced in this article. In particular, it is shown that the use of sliders permits a strong wave localization, tunable by prestress, which opens new routes to metamaterials for mechanical wave control. The fact that a tensile prestress may be applied, leading to in-plane tensile instabilities, implies that the grid analyzed in the present article can effectively be used in practice, because the possibility of out-of-plane buckling is ruled out.

\section{In-plane Floquet–Bloch waves in a grid of preloaded elastic rods equipped with sliders}
\label{sec:formulation_problem}
\begin{figure}[htbp]
    \centering
    \begin{subfigure}{0.45\textwidth}
        \centering        
        \includegraphics[height=0.8\linewidth]{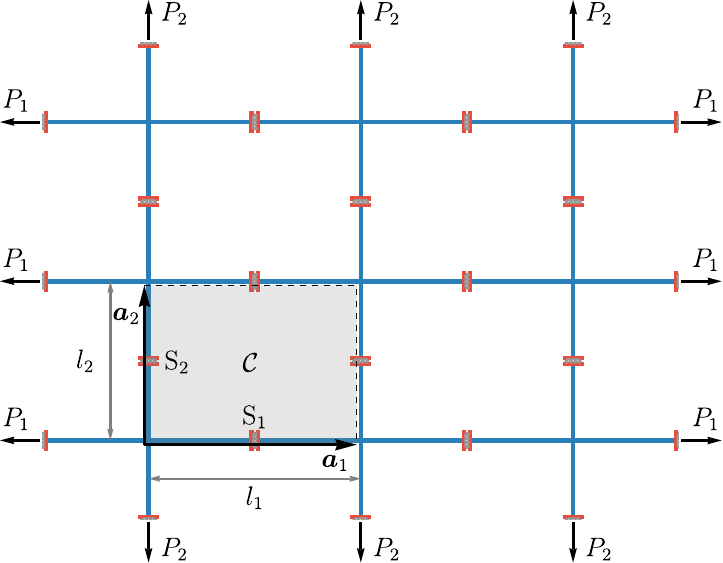}
    \end{subfigure}%
    \hspace{3mm}
    \begin{subfigure}{0.45\textwidth}
        \centering
        \includegraphics[height=0.8\linewidth]{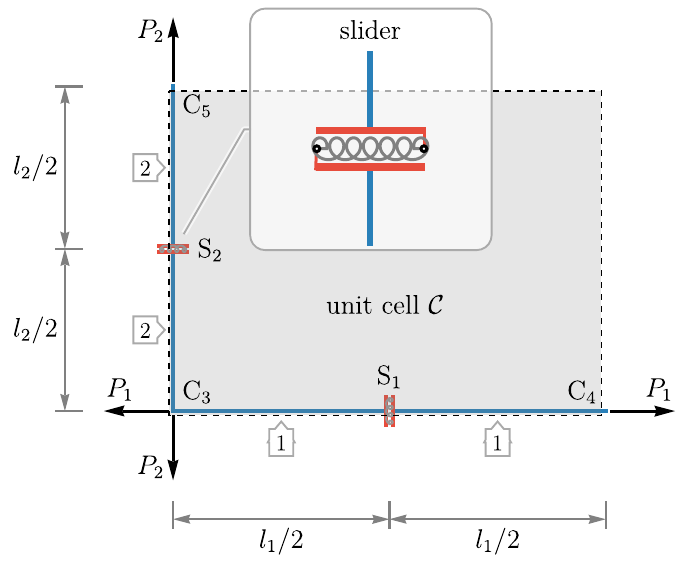}
    \end{subfigure}%
    \caption{
    \label{fig:sliding_grid} Left: a rectangular lattice of axially-preloaded rods, endowed with sliding constraints, realizes an architected material capable of losing ellipticity under both tensile and compressive loadings. Linear springs are introduced to stiffen the sliders, thus preventing the occurrence of \lq floppy modes'. Right: the unit cell with two sliders.
    }
\end{figure}

A two-dimensional periodic lattice of elastic rods is considered, in-plane deformable, both axially and flexurally, in which all structural members are axially preloaded from an unloaded reference configuration, Fig.~\ref{fig:sliding_grid}. 
Two types of nodes connect the rods, one imposing continuity of displacement 
components and rotations (clamped nodes $C_3$, $C_4$, and $C_5$ in the figure), while the other (slider nodes $S_1$ and $S_2$ in the figure) allows a jump in transverse displacement. The latter is linearly related to the shear force transmitted across the node, so that the constraint represents a slider equipped with a linear transverse spring.
The springs inside the sliders are introduced to prevent rigid-body deformation modes (called \lq floppy'). The preload may be produced by tensile or compressive dead loading acting at infinity, while body forces in the lattice are not considered for simplicity. The preload is postulated not only to satisfy equilibrium but also to preserve periodicity and leave the structure initially free from flexure. The incremental response is analyzed by considering arbitrary deformations, which include the development of bending moment and axial and shear forces.

The preloaded configuration, assumed as a reference in an updated Lagrangian description, is periodic along two linearly independent vectors $\{\ba_1,\ba_2\}$, defining the direct basis of the lattice, so that the structure can be constructed from a single unit cell $\mC$, assumed to be composed of $N_{\text{B}}$ (equal to 4 in the present examples, Fig.~\ref{fig:sliding_grid} right part) elastic rods, including Rayleigh rotational inertial effects. Each rod is characterized by length $l$, mass density $\rho$, Young modulus $E$, cross-section area $A$ and second moment of inertia $I$. 
The stiffness of the linear spring embedded inside the sliders is denoted by $k_s$. The equations governing the time-harmonic (in-plane) response with circular frequency $\omega$ of a grid of elastic rods were obtained in \cite{bordiga_2021} and are now complemented with the slider constraints as implemented in \cite{bordiga_tensile_2022}.
The equations governing the time-harmonic dynamics of an elastic unshearable rod are 
\begin{equation}
    \label{eq:governing_equations}
    EA\, \nderiv{u(s)}{s}{2} + \rho A\,\omega^2 u(s) = 0, \qquad
    EI\, \nderiv{v(s)}{s}{4} - \left( P - {\rho I}\, \omega^2 \right) \nderiv{v(s)}{s}{2} - {\rho A}\, \omega^2 v(s) = 0,
\end{equation}
where $P$ is the axial prestress (positive when tensile). Equations \eqref{eq:governing_equations} admit the solutions
\begin{equation}
    \label{eq:u_v_sol}
    u(s) = \sum_{j=1}^2 C_{j}^u\, e^{i\,\beta_j^u\,s} \,, \qquad v(s) = \sum_{j=1}^4 C_{j}^v\, e^{i\,\beta_j^v\,s} \,,
\end{equation}
where $\{C_{1}^u,C_{2}^u,C_{1}^v,...,C_{4}^v\}$ are 6 arbitrary complex constants and the characteristic roots $\beta_j^{u}$ and $\beta_j^{v}$ are given by
\begin{equation*}
    \beta_{1,2}^u = \pm \frac{\Omega}{l} \,, \qquad
    \beta_{1,2,3,4}^v = \pm \frac{1}{l\sqrt{2}}\sqrt{-p + \Omega^2 \pm \sqrt{p^2 + (4\lambda^2 - 2\,p)\Omega^2 + \Omega^4}} \,,
\end{equation*}
\begin{equation}
    \Omega = \omega l \sqrt{\frac{\rho}{E}} \,, \quad
    p = \frac{Pl^2}{EI} \,, \quad
    \lambda =l \sqrt{\frac{A}{I}} \,, 
\end{equation}
where $p$ is a dimensionless measure of the prestress and $\lambda$ is the slenderness of the rod. The constants appearing in equation (\ref{eq:u_v_sol}) can be expressed as functions of the generalized (including rotations) displacements at the ends of the rod.

Reference is made to a periodic material, characterized by a unit cell, where continuity of displacement and rotation are to be imposed at all nodes joining with perfect bonding the elastic rods. 
At the nodes where a slider is present, the continuity conditions only involve axial displacement and rotation. 
In the case shown in Fig.~\ref{fig:sliding_grid} on the right, internal nodes are not present and the rectangular cell is characterized by the dimensionless ratios
\begin{equation}
    \label{eq:dimensionless_groups}
    \lambda_i = l_i \sqrt{\frac{A_i}{I_i}} \,, \quad
    \kappa_i = \frac{k_{s,i}\,l_i^3}{EI_i} \,, \quad
    p_i = \frac{P_il_i^2}{EI_i} \,, \quad 
    \xi = l_2/l_1 \,, \quad 
    \chi = A_2/A_1 \,,  \quad 
    \Omega = \omega l_1 \sqrt{\frac{\rho_1}{E}} \,,
\end{equation}
where the index $i$ identifies the rods aligned parallel to the horizontal, $i=1$, and vertical, $i=2$, directions. Moreover, the boundary nodes are subject to the Floquet-Bloch boundary conditions, based on the notion of Bloch wave vector 
\begin{equation}
    \bk = k_1 \be_1 + k_2 \be_2 \,, 
\end{equation}
so that the generalized (which include rotations) displacements $\bu$ at point $\bx$ satisfy 
\begin{equation}
    \bu(\bx + n_1\ba_1 + n_2\ba_2) = \bu(\bx)\, e^{i\, \bk \cdot (n_1\ba_1 + n_2\ba_2)} \,,
\end{equation}
where $n_1$ and $n_2$ are positive and negative integers (including 0), $\ba_1 = l_1 \be_1$, and $\ba_2 = l_2 \be_2$.

Enforcing the Floquet-Bloch conditions at the boundary nodes of the unit cell leads to an algebraic homogeneous linear system, which governs the wave propagation as the solution of
\begin{equation}
    \label{eq:system}
    \bA(\Omega,\bK,\lambda_1,\lambda_2,\xi,\chi,\kappa_1,\kappa_2,p_1,p_2)\, \bq = \bzero \,,
\end{equation}
where, in the case of the structure shown in Fig.~\ref{fig:sliding_grid}, $\bA$ is a $11 {\times} 11$ complex matrix, function of the dimensionless angular frequency $\Omega$, the dimensionless wave vector $\bK = k_1 l_1 \be_1 + k_2 l_2 \be_2$ and parameters \eqref{eq:dimensionless_groups}.
The vector $\bq$ collects the $11$ degrees of freedom (displacements and rotations of the joints) of the unit cell. Specifically, the vector gathering the degrees of freedom is composed as follows
\begin{equation}
    \begin{aligned}
        & u_1, v_1^{\text{Left}}, v_1^{\text{Right}}, \theta_1, && \text{in the slider S1}\,, \\
        & u_2^{\text{Down}}, u_2^{\text{Up}}, v_2, \theta_2, && \text{in the slider S2}\,, \\
        & u_3, v_3, \theta_3, && \text{in the clamp C3}\,. 
    \end{aligned}
\end{equation}
Furthermore, Floquet-Bloch conditions assume the form
\begin{equation}
    \begin{aligned}
        & \{u_4, v_4, \theta_4\} = \{u_3, v_3, \theta_3\} \exp(i\, \bk \cdot \ba_1), && \text{between nodes C4 and C3}\,, \\
        & \{u_5, v_5, \theta_5\} = \{u_3, v_3, \theta_3\} \exp(i\, \bk \cdot \ba_2), && \text{between nodes C5 and C3}\,. 
    \end{aligned}
\end{equation}
Non-trivial solutions of system \eqref{eq:system} are found when the matrix $\bA$ is rank-deficient
\begin{equation}
    \label{eq:dispersion}
    \det\bA(\Omega,\bK,\lambda_1,\lambda_2,\xi,\chi,\kappa_1,\kappa_2,p_1,p_2) = 0\,,
\end{equation}
which represents the \textit{dispersion equation}, implicitly defining the relation between the angular frequency $\Omega$ and the wave vector $\bK$, the so-called \textit{dispersion relation}.
Furthermore, for each point of the $\{\Omega,\bK\}$-space satisfying Eq.~\eqref{eq:dispersion}, the corresponding eigenvector $\bq(\Omega,\bK)$ can be computed from Eq.~\eqref{eq:system}. Hence, the propagation of Floquet-Bloch waves is governed by the generalized eigenvalue problem~\eqref{eq:system}, where the \textit{eigenfrequencies} are determined by the dispersion relation $\Omega(\bK)$, periodic with period $\left[0,2\pi\right]{\times}\left[0,2\pi\right]$, and the \textit{eigenmodes} (or waveforms) are defined by the eigenvectors $\bq(\Omega,\bK)$.

Dispersion surfaces are provided in the following for the Rayleigh lattices, with an emphasis on the effects of both sliders and preload. To this end, all the geometric parameters of the grid are fixed, and a square lattice, $\xi = 1$, made up of rods of equal characteristics, $\chi = 1$, and $\lambda_1=\lambda_2=15$, is considered. The effect of the slider stiffnesses $\kappa_i$, assumed equal for simplicity $\kappa=\kappa_1=\kappa_2$ and of the preload $p_i$ is investigated.
To highlight the interplay between sliders and preload, the dynamic responses of the lattice with and without preload are separately analyzed.

\section{Lattice without preload}
\subsection{Dispersion properties of Bloch waves} 
\label{sec:dispersion}
The preload is now neglected, $p_1=p_2=0$, so that only the influence of the sliders on the lattice dynamics is investigated. The stiffness of the sliders, assumed to be the same in the horizontal and vertical directions $\kappa=\kappa_1=\kappa_2$, ranges from a low value $\kappa=10^{-2}$ up to an extreme value $\kappa=10^{8}$, for which the response of the lattice reduces to that of a perfectly jointed square lattice, in other words without sliders, analyzed in \cite{bordiga_tensile_2022}.
%
\begin{figure}[htb!]
    \centering
    \begin{subfigure}{0.33\textwidth}
        \centering
        \includegraphics[width=0.95\linewidth]{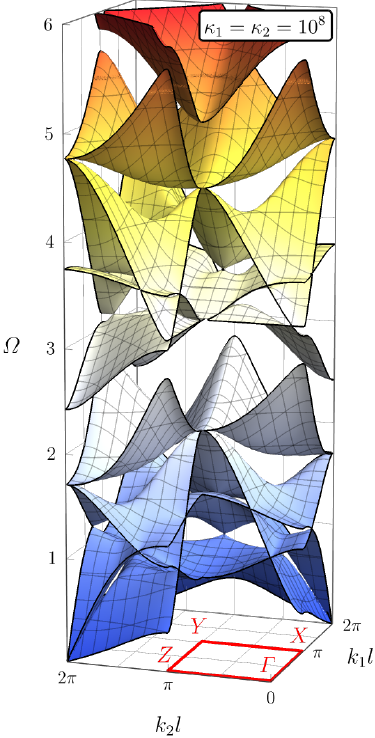}
    \end{subfigure}%
    \begin{subfigure}{0.33\textwidth}
        \centering
        \includegraphics[width=0.95\linewidth]{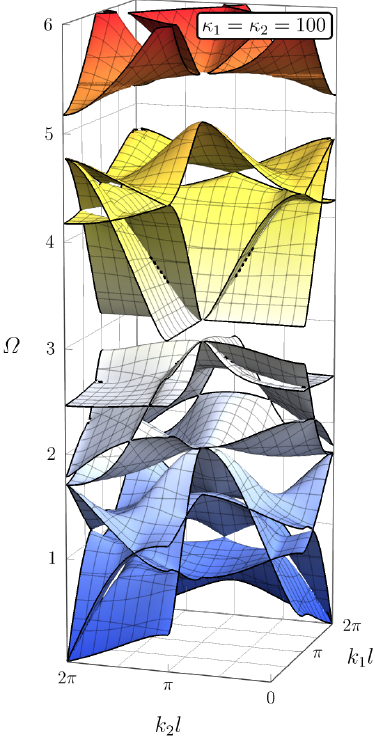}
    \end{subfigure}%
    \begin{subfigure}{0.33\textwidth}
        \centering
        \includegraphics[width=0.95\linewidth]{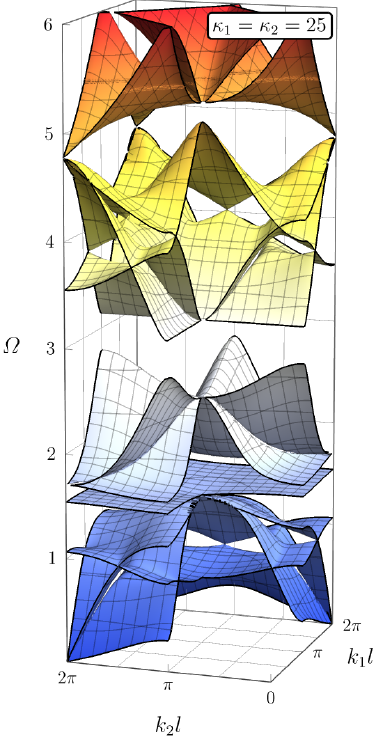}
    \end{subfigure} \\[5mm]
    \begin{subfigure}{0.33\textwidth}
        \centering 
        \includegraphics[width=0.95\linewidth]{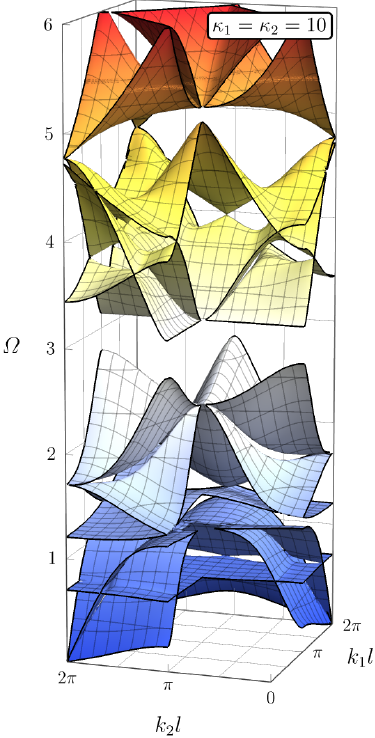}
    \end{subfigure}%
    \begin{subfigure}{0.33\textwidth}
        \centering
        \includegraphics[width=0.95\linewidth]{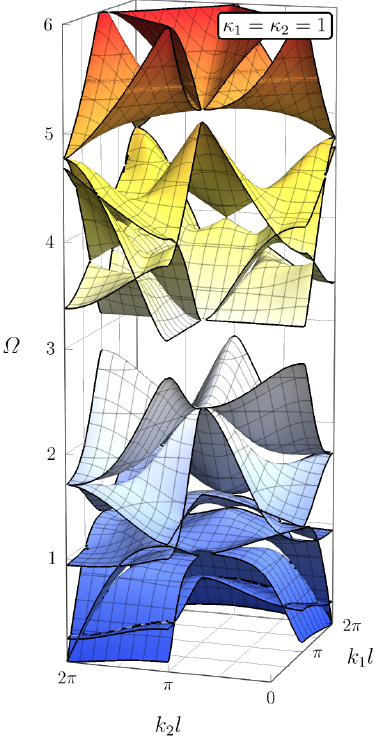}
    \end{subfigure}%
    \begin{subfigure}{0.33\textwidth}
        \centering 
        \includegraphics[width=0.95\linewidth]{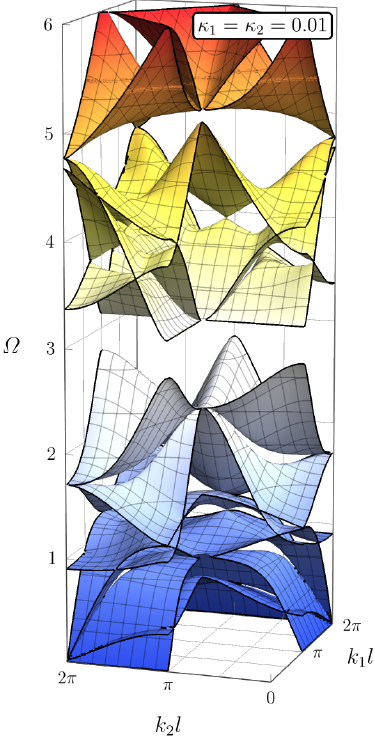}
    \end{subfigure}
    \caption{
    \label{fig:surfcut} Dispersion surfaces of a squared grid of Rayleigh rods with slenderness $\lambda =15$, without preload. Different values of the stiffness of the sliders are used $\kappa=10^{8},100,25,10,1,10^{-2}$. 
    The case $\kappa=10^{8}$ is reported as a reference because the stiffness of the sliders is so high that they do not have any effect and the behaviour of the grid is the same as in the absence of sliders. 
    The path $\Gamma$--$X$--$Y$--$Z$--$\Gamma$ reported in red represents the first Brillouin zone. In the following dispersion curves are plotted along the triangular path $\Gamma$--$X$--$Y$--$\Gamma$ (for the lattice without preload, which possesses square symmetry) or along the squared path $\Gamma$--$X$--$Y$--$Z$--$\Gamma$ (for the lattice with unequal preload, which possesses orthotropic symmetry).
    }
\end{figure}

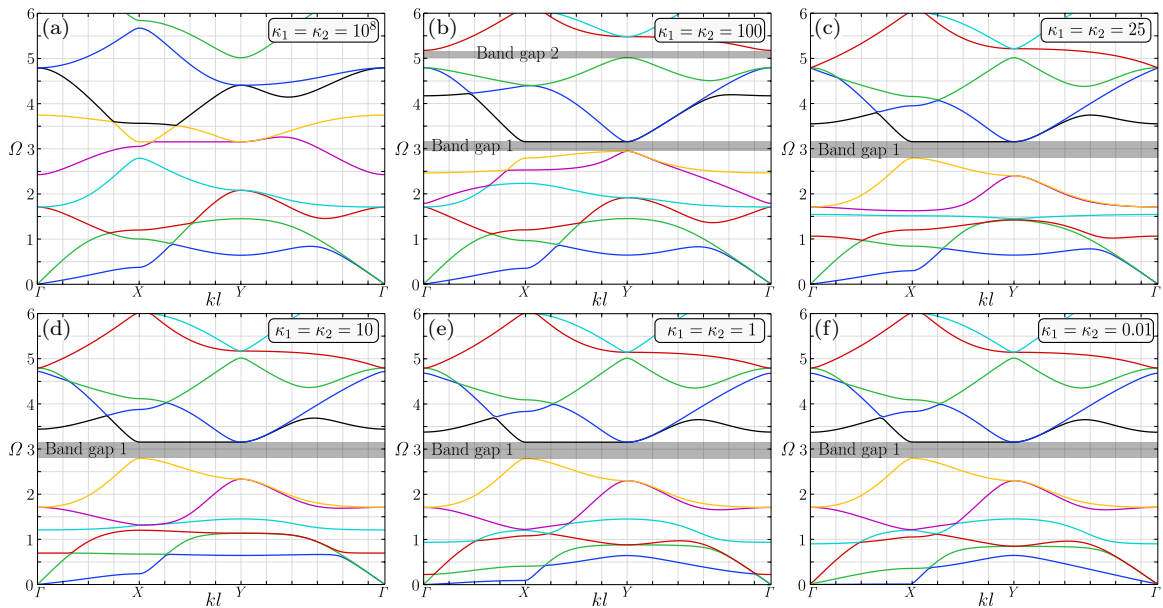
\begin{figure}[htb!]
    \centering
    \begin{subfigure}{0.33\textwidth}
        \centering
        {\phantomsubcaption\label{fig:Brillouin_100000000}}
        \scalefont{0.8}
        \begin{tikzpicture}
        \node[inner sep=0pt,anchor=north west] (figa) at (0,0)
        {\includegraphics[width=0.98\linewidth]{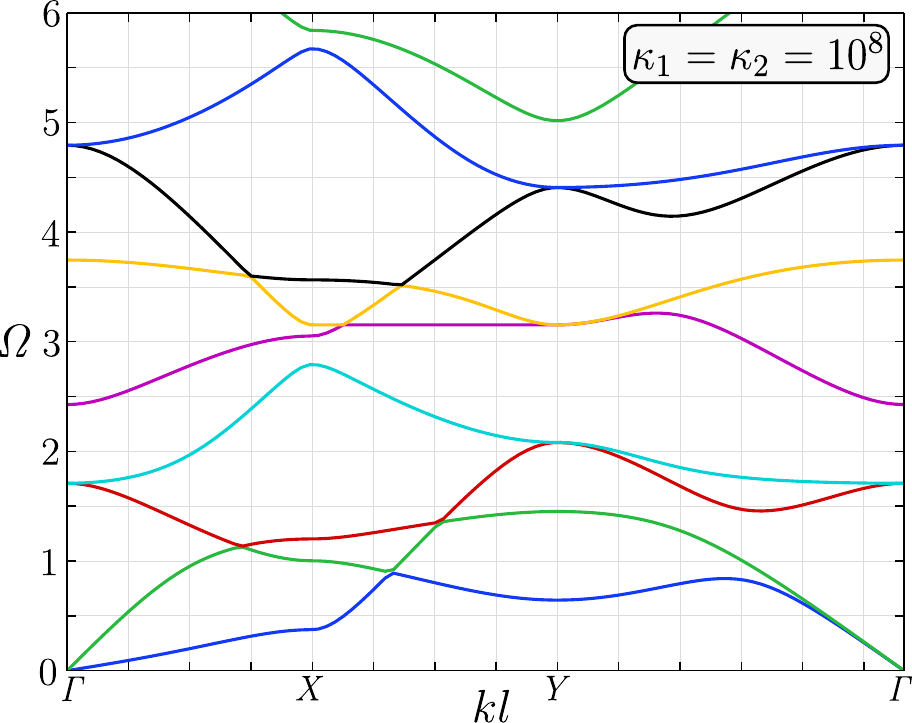}};
        \node[inner sep=0pt,align=center,fill=white,rounded corners=1pt,anchor=north west] at (0.42,-0.14) {(a)};
        \end{tikzpicture}
    \end{subfigure}%
    \begin{subfigure}{0.33\textwidth}
        \centering
        {\phantomsubcaption\label{fig:Brillouin_100}}
        \scalefont{0.8}
        \begin{tikzpicture}
        \node[inner sep=0pt,anchor=north west] (figa) at (0,0)
        {\includegraphics[width=0.98\linewidth]{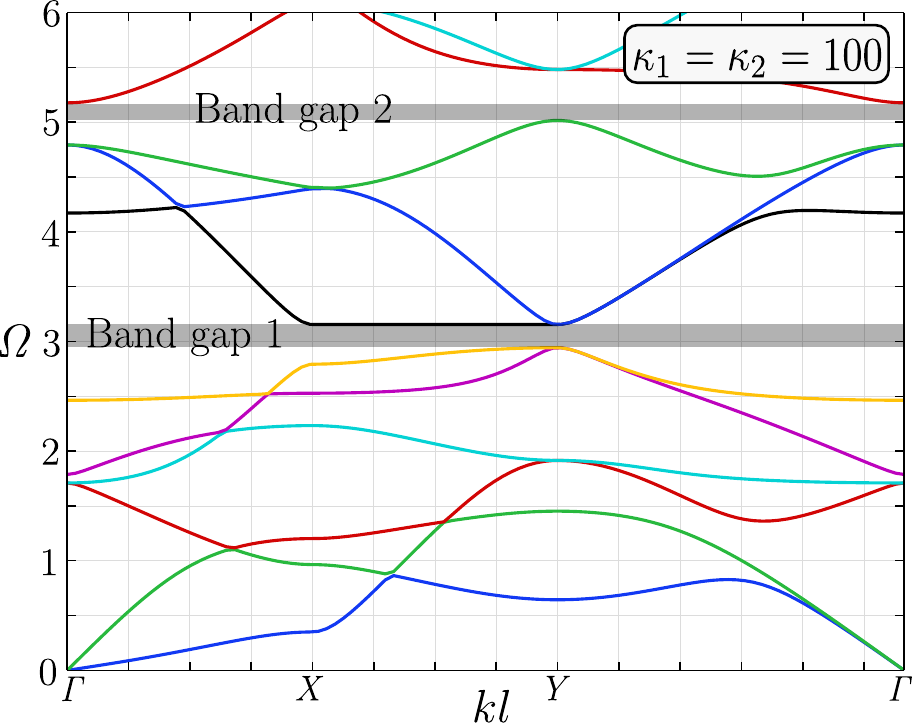}};
        \node[inner sep=0pt,align=center,fill=white,rounded corners=1pt,anchor=north west] at (0.42,-0.14) {(b)};
        \end{tikzpicture}
    \end{subfigure}%
    \begin{subfigure}{0.33\textwidth}
        \centering
        {\phantomsubcaption\label{fig:Brillouin_25}}
        \scalefont{0.8}
        \begin{tikzpicture}
        \node[inner sep=0pt,anchor=north west] (figa) at (0,0)
        {\includegraphics[width=0.98\linewidth]{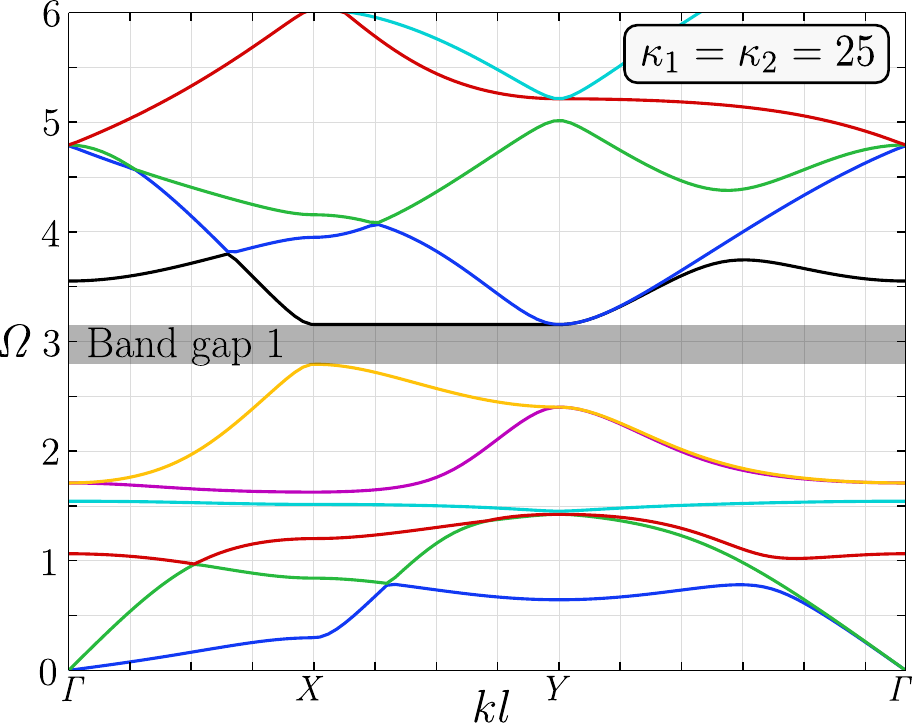}};
        \node[inner sep=0pt,align=center,fill=white,rounded corners=1pt,anchor=north west] at (0.42,-0.14) {(c)};
        \end{tikzpicture}
    \end{subfigure}\\
    \begin{subfigure}{0.33\textwidth}
        \centering
        {\phantomsubcaption\label{fig:Brillouin_10}}
        \scalefont{0.8}
        \begin{tikzpicture}
        \node[inner sep=0pt,anchor=north west] (figa) at (0,0)
        {\includegraphics[width=0.98\linewidth]{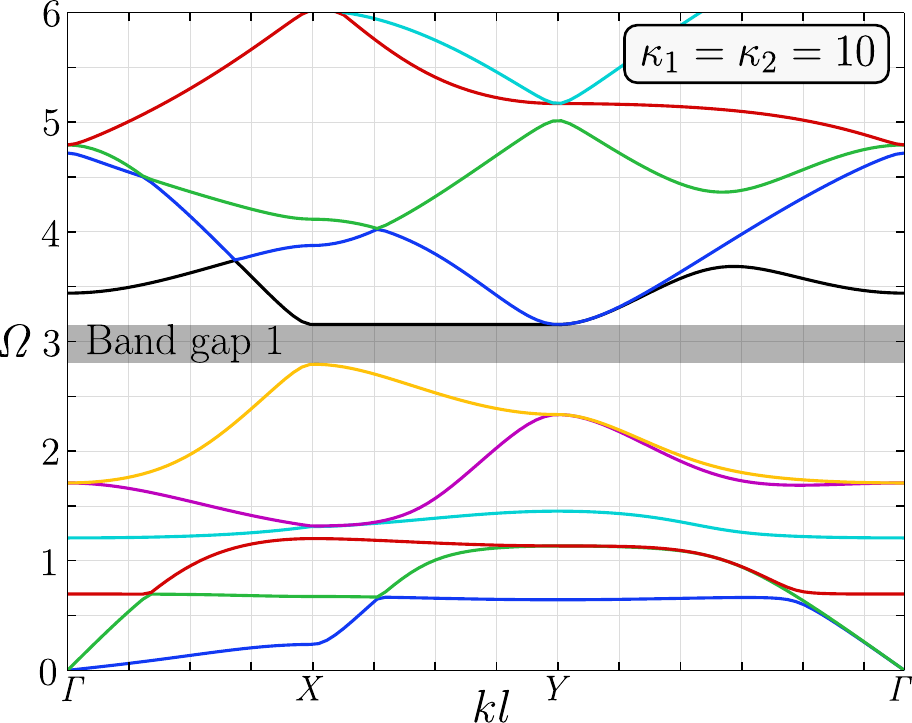}};
        \node[inner sep=0pt,align=center,fill=white,rounded corners=1pt,anchor=north west] at (0.42,-0.14) {(d)};
        \end{tikzpicture}
    \end{subfigure}
    \begin{subfigure}{0.33\textwidth}
        \centering
        {\phantomsubcaption\label{fig:Brillouin_1}}
        \scalefont{0.8}
        \begin{tikzpicture}
        \node[inner sep=0pt,anchor=north west] (figa) at (0,0)
        {\includegraphics[width=0.98\linewidth]{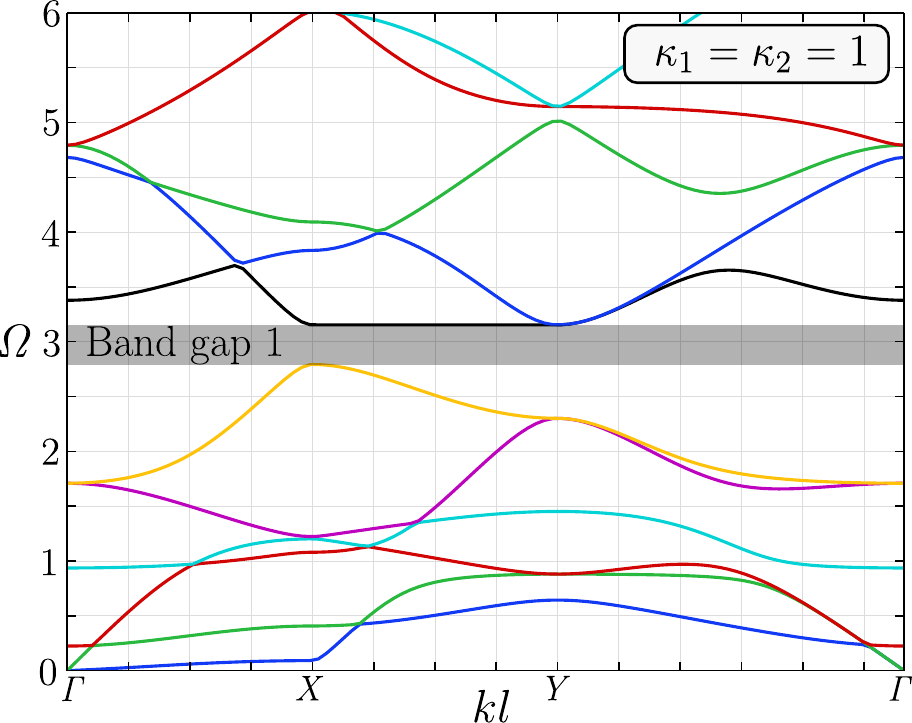}};
        \node[inner sep=0pt,align=center,fill=white,rounded corners=1pt,anchor=north west] at (0.42,-0.14) {(e)};
        \end{tikzpicture}
    \end{subfigure}
    \begin{subfigure}{0.33\textwidth}
        \centering
        {\phantomsubcaption\label{fig:Brillouin_001}}
        \scalefont{0.8}
        \begin{tikzpicture}
        \node[inner sep=0pt,anchor=north west] (figa) at (0,0)
        {\includegraphics[width=0.98\linewidth]{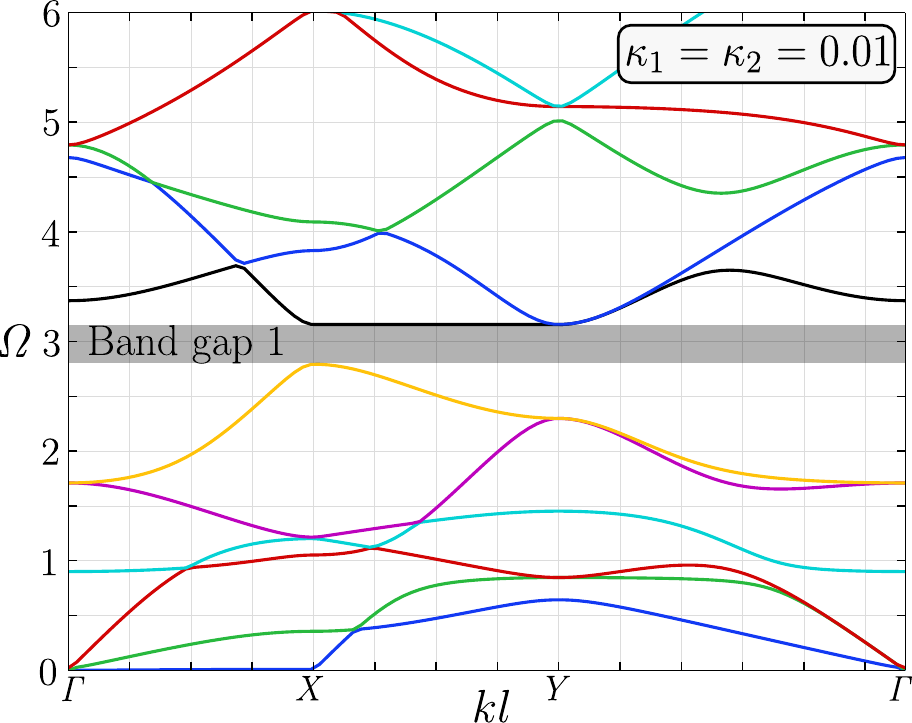}};
        \node[inner sep=0pt,align=center,fill=white,rounded corners=1pt,anchor=north west] at (0.42,-0.14) {(f)};
        \end{tikzpicture}
    \end{subfigure}%
    \caption{
    \label{fig:disp_diagram} Dispersion curves for a square grid of elastic rods without preload, for different values of the slider stiffness  $\kappa=10^{8},100,25,10,1,10^{-2}$. The diagrams have been evaluated along the boundary of the first irreducible Brillouin zone (path $\Gamma$--$X$--$Y$--$\Gamma$ sketched in Fig.~\ref{fig:surfcut}). The relative band gaps are highlighted in figures \subref{fig:Brillouin_100}-\subref{fig:Brillouin_001}. \lq Twin band gaps' are visible in panel \subref{fig:Brillouin_100}.
    }
\end{figure}

Dispersion surfaces are shown in Fig.~\ref{fig:surfcut} and complemented by the band diagrams reported in Fig.~\ref{fig:disp_diagram}, relative to the path $\Gamma$--$X$--$Y$--$\Gamma$ shown in the former figure. The latter figure allows the appreciation of details that would remain undetected from the dispersion surfaces. 
In particular, the result reported in Fig.~\ref{fig:Brillouin_100000000}, pertaining to the case in which the slider stiffness is so high that the rods result fully connected  (case already analyzed in \cite{bordiga_tensile_2022}), is used as reference. The dispersion diagrams show that, at sufficiently high slider stiffness, the presence of the sliders only affects the vibrational response of the lattice in the high-frequency regime, whereas the acoustic curves result weakly altered. Conversely, changes are limited to low frequencies $\Omega$ at small stiffness of the sliders. In this case, at low frequencies, the reduction of the slider stiffness induces a decrease in the slope of the acoustic branches. This is pronounced for the acoustic curve that characterizes the propagation of shear waves, while pressure waves are less influenced by a change in the slider stiffness. For extremely low values of $\kappa$, the slope of the acoustic curve, responsible for the propagation of the shear waves, tends to zero along the two axes $K_1$ and $K_2$, while the dispersion surface does not touch zero everywhere, as shown in Fig.~\ref{fig:surfcut}. 
Another peculiarity of the elastic grid is the lowering of the second and third dispersion surfaces as $\kappa$ is reduced. When the slider stiffness tends to zero, the third dispersion surface touches the origin, so that now three dispersion surfaces emanate from the origin, instead of the usual two.

An infinite set of standing waves propagating at the same frequency with an arbitrary wave vector can be produced through the tuning of the stiffness of the sliders. Indeed, for a stiffness $\kappa=10$, an almost flat surface becomes visible in Fig.~\ref{fig:Brillouin_10}, induced by the flattening of the third dispersion surface.  Another interesting effect introduced by the sliders can be noticed at high frequency, where an opening of two band gaps is observed, Fig.~\ref{fig:Brillouin_100}.

Band gaps are not present in the absence of sliders (Fig.~\ref{fig:Brillouin_100000000}), but when the slider stiffness reduces, a first band gap, around the frequency $\Omega=3$, appears and remains almost unaltered as $\kappa$ tends to zero. A second band gap, visible around $\Omega=5$, appears, so that \lq twin band gaps' are present. The second band gap enlarges as the slider stiffness reduces until a critical value is reached, after which this band gap begins to shrink and eventually disappears, Fig.~\ref{fig:Brillouin_25}--\subref{fig:Brillouin_001}.
%

\FloatBarrier

\subsection{Forced vibration}
\label{sec:forced_vibration}
The relation between the dynamic response of a grid of Rayleigh rods and the Floquet-Bloch analysis performed in the previous section can be investigated through the analysis of the vibrations induced by a time-harmonic source (a concentrated force or moment) in a lattice of infinite extent.
To this purpose, a grid of rods is numerically analyzed using the Comsol Multiphysics$^{\circledR}$ F.E. program in the frequency response mode.
A square finite-size computational window with $(N-1) {\times} (N-1)$ unit cells is considered, where $N=161$ is the number of nodes assumed in each direction. The window is bounded with a perfectly matched layer (PML), to simulate the response of an infinite lattice. In fact, by tuning the damping in the boundary layers, the outgoing waves can be completely absorbed, so that reflection is not generated in the interior domain.

A concentrated time-harmonic source is applied (in-plane) to the central junction ($C_3$ in Fig.~\ref{fig:sliding_grid}), pulsating with a dimensionless angular frequency $\Omega$. For a given loading and a given dimensionless angular frequency $\Omega$, the complex displacement field, with horizontal and vertical components 
\begin{equation}
    u=u_R+iu_I, \quad v=v_R+iv_I,     
\end{equation}
is computed. The results are plotted in terms of the displacement amplitude associated with the real parts, 
\begin{equation}
    \delta_{R}(x,y,\Omega) = \sqrt{u_R^2+v_R^2}\,,     
\end{equation}
while the displacement amplitude associated with the imaginary parts, 
\begin{equation}
    \delta_{I}(x,y,\Omega) = \sqrt{u_I^2+v_I^2}\,,    
\end{equation}
is omitted for conciseness. 
\vspace{0.5cm}

\begin{figure}[htb!]
    \centering
    \begin{subfigure}{0.33\textwidth}
        \centering
        \includegraphics[width=0.98\linewidth]{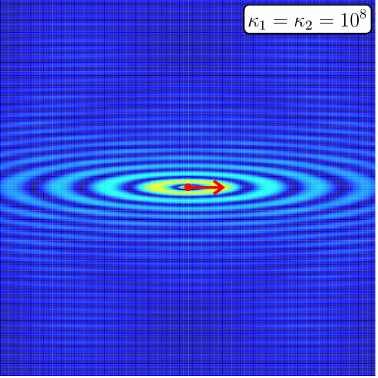}
    \end{subfigure}%
    \begin{subfigure}{0.33\textwidth}
        \centering
        \label{fig:Forc_grid_1}
        \includegraphics[width=0.98\linewidth]{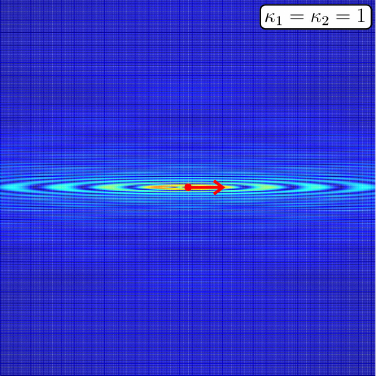}
    \end{subfigure}%
    \begin{subfigure}{0.33\textwidth}
        \centering
        \includegraphics[width=0.98\linewidth]{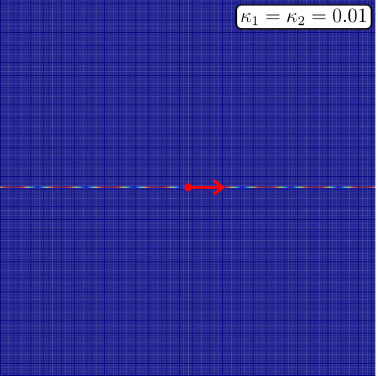}
    \end{subfigure}
    \caption{
    \label{fig:LowRegime} Maps of displacement amplitude (obtained through f.e. analyses) for different stiffness of the sliders $\kappa=\{10^8,1,10^{-2}\}$ of a grid of Rayleigh rods (without preload). The latter is excited by a time-harmonic concentrated force (aligned parallel to the horizontal direction) vibrating at low frequency, $\Omega=0.05$. When the slider stiffness becomes vanishing small a strong localization of the signal is observed, visible on the right.
    }
\end{figure}

\paragraph{Low frequency regime}
\FloatBarrier
The elastic grid is forced by a horizontal force, pulsating at the low angular frequency $\Omega=0.05$. 
Results in terms of displacement amplitude $\delta_{R}$ are reported in Fig.~\ref{fig:LowRegime}, where different values of the slider stiffness $\kappa$ are investigated. 
%
%

At these frequencies, two dispersion surfaces are always intersected. In the long-wavelength regime, at a very low value of the slider stiffness, $\kappa=10^{-2}$ Fig.~\ref{fig:LowRegime} on the right, the dynamic response exhibits a strong localization along the directions of the applied load, because the compliance of the sliders does not allow the propagation of waves through the adjacent (horizontal) rows of rods. In other words, the horizontal wave localization can be viewed in terms of the response of the elastic material equivalent to the grid which loses ellipticity when the stiffness of the sliders is approaching zero. 

This transmission becomes possible when the slider stiffness is increased, so that shear waves are allowed to propagate throughout the whole lattice, Fig.~\ref{fig:LowRegime} on the centre. At the highest value of slider stiffness, the rods are perfectly jointed, so that the response of a square lattice without sliders is obtained, Fig.~\ref{fig:LowRegime} on the left. The growth of the wavelength with the increase of the slider stiffness can be deduced from the dispersion curves reported in Fig.~\ref{fig:disp_diagram}, where high values of the slider stiffness increase the slope of the acoustic curve, responsible for shear wave propagation. Therefore, the modulus of the wave vector $\bK$ reduces, and consequently, wave fronts are more distant.

\FloatBarrier
\vspace{0.5cm}

\begin{figure}[htb!]
    \centering
    \begin{subfigure}{0.33\textwidth}
        \centering
        \includegraphics[width=0.98\linewidth]{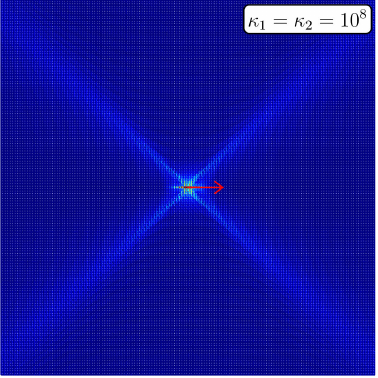}
    \end{subfigure}%
    \begin{subfigure}{0.33\textwidth}
        \centering
    \includegraphics[width=0.98\linewidth]{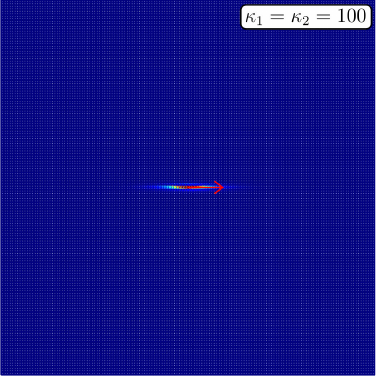}
    \end{subfigure}%
    \begin{subfigure}{0.33\textwidth}
        \centering
        \includegraphics[width=0.98\linewidth]{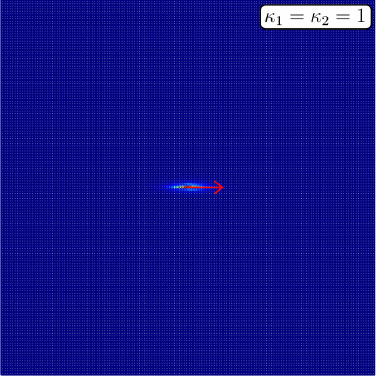}
    \end{subfigure}
    \caption{
    \label{fig:bandgap1} As for Fig.~\ref{fig:LowRegime}, except that the frequency $\Omega=3.0$ is investigated for $\kappa=\{10^8,100,1\}$. The selected frequency falls within the low-frequency band gap of Fig.~\ref{fig:disp_diagram} for $\kappa=100$ and 1, so propagation is here suppressed.
    }
\end{figure}

\begin{figure}[htb!]
    \centering
    \begin{subfigure}{0.33\textwidth}
        \centering
        \includegraphics[width=0.98\linewidth]{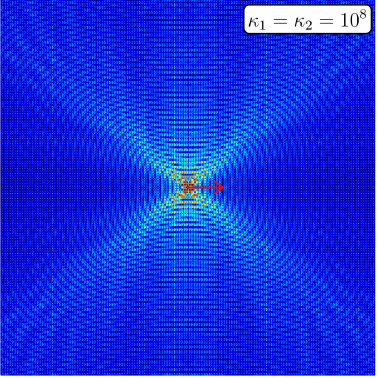}
    \end{subfigure}%
    \begin{subfigure}{0.33\textwidth}
        \centering
        \includegraphics[width=0.98\linewidth]{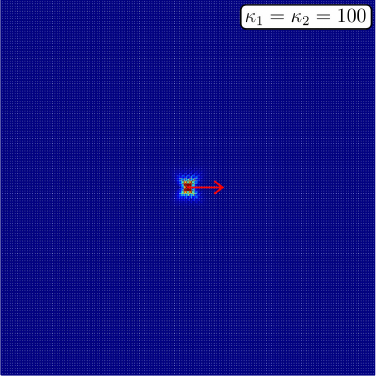}
    \end{subfigure}%
    \begin{subfigure}{0.33\textwidth}
        \centering
        \includegraphics[width=0.98\linewidth]{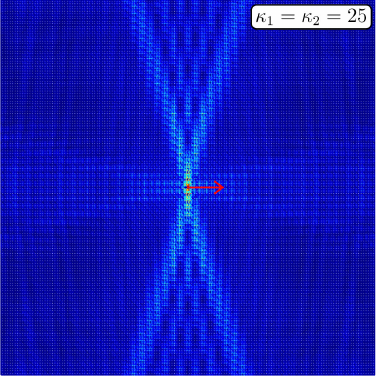}
    \end{subfigure}
    \caption{
    \label{fig:bandgap2} As for Fig.~\ref{fig:bandgap1}, except that the vibration frequency, $\Omega=5.1$, is internal to the high frequency band gap visible in Fig.~\ref{fig:disp_diagram} at $\kappa=100$. Here propagation is suppressed. Decreasing the stiffeness of the sliders to $\kappa=25$, the high-frequency-band gap closes and propagation is again possible.
    }
\end{figure}

\paragraph{Band gaps}
An analysis of the dispersion curves in Figs.~\ref{fig:disp_diagram} reveals that twin band gaps are present in the range of analyzed frequencies. The first band gap, at low frequency, is not present when the rods result perfectly jointed ($\kappa=10^8$), but it appears when the stiffness of the sliders is reduced, $\kappa =100$ and 1. 
The response of the lattice is shown in Fig.~\ref{fig:bandgap1} at the frequency $\Omega=3$ and in Fig.~\ref{fig:bandgap2} at the frequency $\Omega=5.1$. The former frequency is internal to the band gap at low frequency for $\kappa= \{100,1\}$, while the latter is internal to the band gap at high frequency, for $\kappa=100$. The perfectly jointed rod case is also reported for comparison, $\kappa=10^8$, where no band gap is present. The figure shows that the behaviour of the forced lattice confirms the dynamic response predicted by the dispersion curves. 
%
%
%
Differently from the first band gap, the latter is present for a limited interval of values of the slider stiffness. The band gap vanishes for $\kappa=25$, as visible in the right panel of Fig.~\ref{fig:bandgap2}.

\paragraph{\lq Quasi-overdetermined' lattice}
In the limit, when the slider stiffness vanishes, the grid is susceptible to non-trivial rigid-body motions (translations aligned parallel to the sliders), so that the system of equations governing the statics of the grid becomes overdetermined. 
The analysis of a case near this limit, in which the slider stiffness is very low, $\kappa=0.01$, is interesting. It will be referred to as \lq quasi-overdetermined' and can be viewed as a loss of ellipticity at vanishing slider stiffness of the equivalent elastic material obtained through homogenization. 

\begin{figure}[htb!]
    \centering
    \begin{subfigure}{0.33\textwidth}
        \centering
        \includegraphics[width=0.98\linewidth]{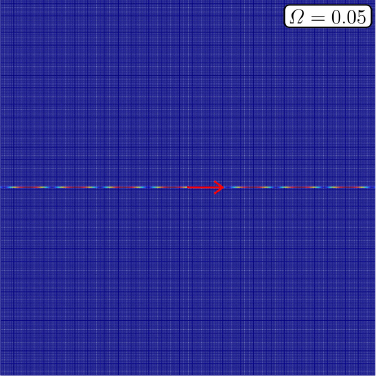}
    \end{subfigure}%
    \begin{subfigure}{0.33\textwidth}
        \centering
        \includegraphics[width=0.98\linewidth]{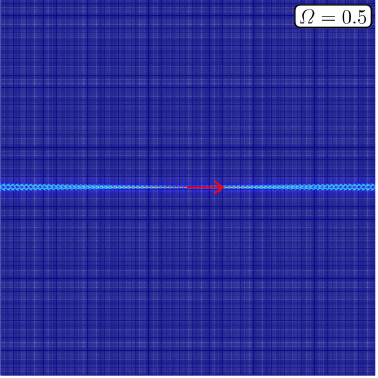}
    \end{subfigure}%
    \begin{subfigure}{0.33\textwidth}
        \centering
        \includegraphics[width=0.98\linewidth]{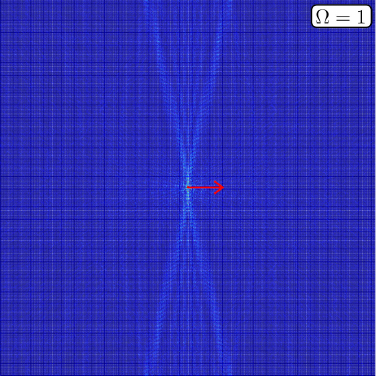}
    \end{subfigure}\\
    \caption{
    \label{fig:ForceP0labile} Vibrations of a \lq quasi-overdetermined' grid of Rayleigh rods, characterized by a low slider stiffness $\kappa_1=\kappa_2=0.01$, in the absence of preload and excited by a time-harmonic horizontal force pulsating at different frequencies. The displacement field is strongly localized at low frequencies $\Omega=0.05$ and $0.5$, as the effect of the low compliance of the sliders, while at high frequency, $\Omega = 1$, the wave spreads to the entire lattice.
    }
\end{figure}

\begin{figure}[htb!]
    \centering
    \begin{subfigure}{0.33\textwidth}
        \centering
        \includegraphics[width=0.99\linewidth]{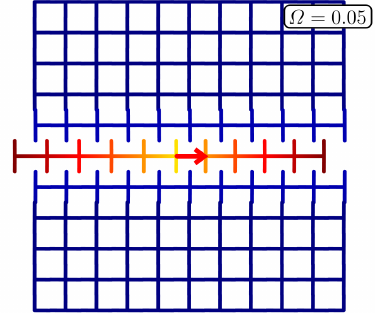}
    \end{subfigure}%
    \begin{subfigure}{0.33\textwidth}
        \centering
        \includegraphics[width=0.99\linewidth]{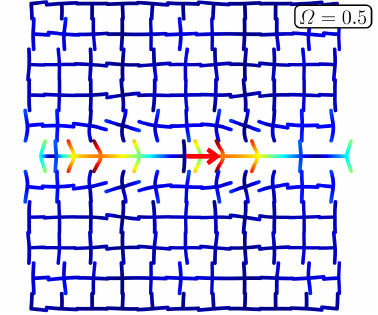}
    \end{subfigure}%
    \begin{subfigure}{0.33\textwidth}
        \centering
        \includegraphics[width=0.99\linewidth]{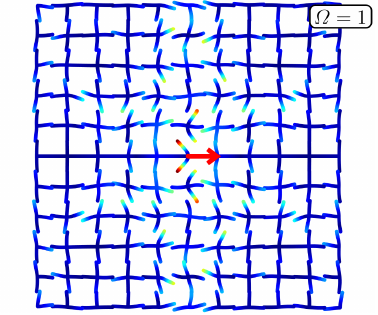}
    \end{subfigure}\\
    \caption{
    \label{fig:ForceP0zoom} Details of Fig.~\ref{fig:ForceP0labile} showing localization of the vibrational displacement, in the proximity of the row where the load is applied at $\Omega=0.05$ and 0.5, while at high frequency $\Omega=1$, waves spread through the \lq quasi-overdetermined' grid of rods, characterized by small slider stiffness $\kappa_1=\kappa_2=0.01$.
    }
\end{figure}

When their stiffness becomes low, sliders can transmit with difficulty a signal when the lattice is forced in a direction aligned parallel to them, as confirmed in Fig.~\ref{fig:ForceP0labile} for $\Omega=0.05$ and $0.5$, where the propagation is highly localized. This is typical of the low-frequency regime, close to the origin $\Omega=0$. Differently, at sufficiently high frequency, $\Omega=1$, the propagation of a signal is not excluded, as foreseen by the dispersion curves. 
The situation is also illustrated in Fig.~\ref{fig:ForceP0zoom}, which reports details of Fig.~\ref{fig:ForceP0labile} (so magnified that the grid of rods becomes visible) cut near the point of application of the pulsatile force. 

Fig.~\ref{fig:ForceP0zoom} shows the influence of the inertia on the dynamic response of the lattice. A horizontal translation of the row where the load is applied and of the connected vertical rods is visible for $\Omega=0.05$, where bending is negligible, so that the rotation transmitted by the sliders is small. Consequently, the signal is strongly localized. Increasing the frequency to $\Omega=0.5$ and $\Omega=1$, the inertia of the rods induces their bending, so the rotation of the sliders propagates the signal in the whole grid.

\begin{figure}[htb!]
    \centering
    \begin{subfigure}{0.33\textwidth}
        \centering
        \includegraphics[width=0.98\linewidth]{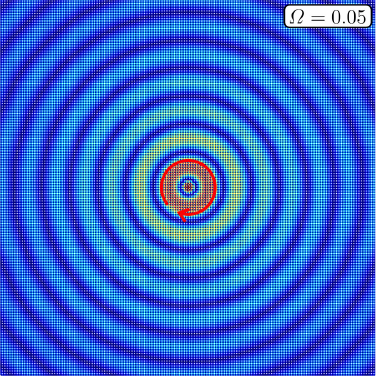}
    \end{subfigure}%
    \begin{subfigure}{0.33\textwidth}
        \centering
        \includegraphics[width=0.98\linewidth]{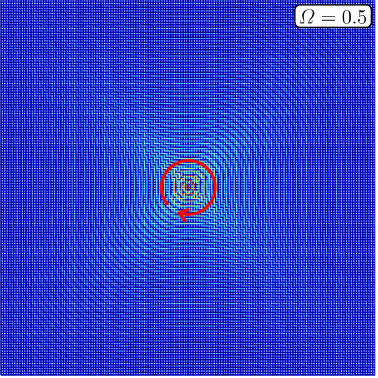}
    \end{subfigure}%
    \begin{subfigure}{0.33\textwidth}
        \centering
        \includegraphics[width=0.98\linewidth]{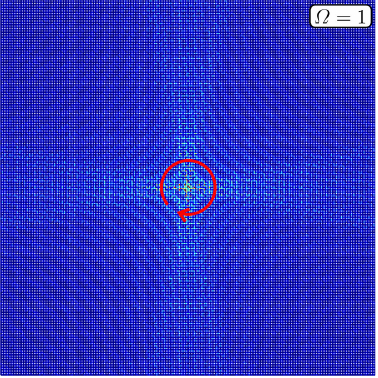}
    \end{subfigure}%
    \caption{
    \label{fig:MomentP0} A concentrated moment applied at the centre of a \lq quasi-overdetermined' grid, $\kappa_1=\kappa_2=0.01$, (without preload) leads to propagation in the whole lattice at all frequencies. The effects of an applied pulsatile moment contrast with those related to the application of a pulsatile force, see Fig.~\ref{fig:ForceP0labile}. In fact, now waves propagate in every direction.
    }
\end{figure}

The response of the grid to a concentrated moment (applied at the central node of the grid) is investigated in Fig.~\ref{fig:MomentP0}. 
In this case, bending of the rods is induced at all frequencies, even low ones, so that a circular pattern is observed. This pattern is present until $\Omega=0.5$ is reached, while at higher frequencies the propagation of the signal changes, so that it becomes oriented almost parallel to the axes of the grid.
%
%
The behaviour of the grid under an applied moment is in stark contrast with the applied force (see  Figs.~\ref{fig:ForceP0labile} and \ref{fig:ForceP0zoom}) because the moment excites waves propagating in every direction.

\section{Preloaded lattice: dispersion properties, forced vibration and ‘macro’ bifurcation}
\label{sec:Prestress}
The investigation of the stability of the grid includes the preload of the rods. Lattice bifurcations can exhibit deformation modes in the grid of rods with various wavelengths, and they are controlled by the value of the preload state. A \lq global' or \lq macro' bifurcation occurs in the limit of infinite wavelength; otherwise, the bifurcation is referred to as \lq microscopic'. Only the macro (or global) instability can be associated with the failure of ellipticity of the equivalent elastic prestressed solid, which involves the appearance of an abrupt localization of the deformation within the lattice.
The latter is known as \lq shear band mode' and represents in the lattice the analogue of a jump in the incremental strain in the equivalent solid. It is important to note how the tunability of the slider stiffness and preload enables the characteristics of the grid to control the occurrence of shear bands. 

The analyses which follow are complemented by the relations between the grid and its homogenized response in terms of an elastic equivalent continuum. 
The homogenization scheme was derived in \cite{bordiga_tensile_2022, bordiga_2021}  and is not repeated here. It was shown in \cite{bordiga_tensile_2022} that the incremental response of the preloaded grid of elastic rods can be homogenized to become the incremental response of a prestressed elastic solid.

The preload of components $p_1$ and $p_2$ is transformed into a Cauchy prestress of components $T_{11}$ and $T_{22}$,  
\begin{equation}
    T_{11} = \frac{P_1}{l_2} = \frac{EA_1}{l_1} \frac{p_1}{\lambda_1^2\xi}, \quad
    T_{22} = \frac{P_2}{l_1} = \frac{EA_1}{l_1} \frac{p_2\chi}{\lambda_2^2}, 
\end{equation}
affecting the response of an equivalent elastic material defined by a fourth-order tensor, function of the slider stiffness, the grid stiffness and its geometry. 

A macro bifurcation in the grid corresponds to a loss of ellipticity in the continuum, while micro bifurcations leave the equivalent solid unaffected.  
When the stability domain of the grid does not include micro bifurcations, the correspondence between the grid and the equivalent continuum becomes perfect.

Failure of ellipticity occurs in a solid when the speed of an acceleration wave vanishes, a condition represented by an eigenvalue of the acoustic tensor becoming null. The direction $\bn_E$ for which the acoustic tensor (reported in Appendix~\ref{sec:effective_acoustic_tensor} for the case under study) becomes singular, and the respective eigenvector $\bg_E$ represents the normal to the shear band and the shear deformation mode, respectively. 

In the following, two case studies are provided: uniaxial preload in the horizontal direction $p_2=0$ and equibiaxial preload, $p_1=p_2$.

\subsection{Uniaxial preload $p_2 = 0$}
The lattice is subject to an uniaxial preload defined by the axial forces acting in the horizontal direction ($p_2=0$).

The stability domain inside which global bifurcations do not occur is obtained in the $p_1$--$p_2$ space as shown in \cite{bordiga_tensile_2022}. The grid and the preload satisfy orthotropy and in the case under analysis orthotropy reduces to cubic symmetry when the grid is unloaded.   
Therefore, the stability domain is symmetric with respect to the equibiaxial loading axis, $p_1=p_2$ (analyzed in the next subsection). 

Macro bifurcations, corresponding to the formation of shear band modes appear on the boundary of the domain, aligned parallel (orthogonal) to the loading direction for tension (for compression). Due to the presence of the sliders, the stability domain of the lattice is bounded both in compression and in tension and tends to become unbounded in tension when the stiffness of the sliders increases.

For three distinct values of slider stiffness $\kappa$, the preload component $p_1$, which causes a macro bifurcation, is listed in Table~\ref{tab:Critical_Prestress2}, with $p_{1E}^-$ and $p_{1E}^+$ denoting the negative and positive critical preloads, respectively. 
%
\begin{table}[ht]
    \centering
    \begin{tabular}{@{}lcc@{}}
        \toprule
        \multicolumn{3}{c}{Critical values of uniaxial preload $p_2=0$ for macro bifurcation} \\
        \midrule
        Slider stiffness $\kappa$ & Compressive critical prestress $p_{1E}^-$ & Tensile critical prestress $p_{1E}^+$ \\
        \cmidrule(r){1-1} \cmidrule(rl){2-2} \cmidrule(l){3-3}
        1   & $-$0.538088 &  0.929147 \\
        10  & $-$2.66810  &  6.65908  \\
        100 & $-$5.01017  & 34.06669  \\
        \bottomrule
    \end{tabular}
    \caption{
    \label{tab:Critical_Prestress2} Macro bifurcation occurring in the elastic grid for different slider stiffness $\kappa$. Uniaxial preload of the rods is assumed, $p_2=0$.The occurrence of macro bifurcation in the grid coincides with the loss of ellipticity in the equivalent continuum, where preload becomes equivalent to prestress.
    }
\end{table}

\begin{figure}[htb!]
    \centering
    \begin{subfigure}{0.32\textwidth}
        \centering
        \includegraphics[width=0.99\linewidth]{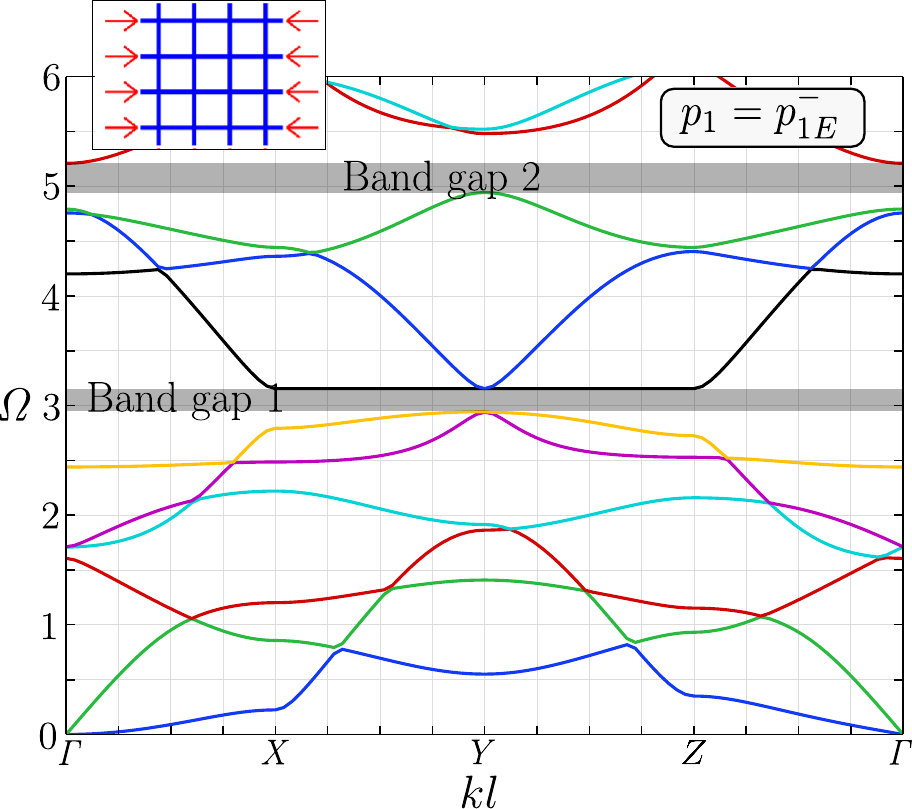}
    \end{subfigure}
    \begin{subfigure}{0.32\textwidth}
        \centering        
        \includegraphics[width=0.99\linewidth]{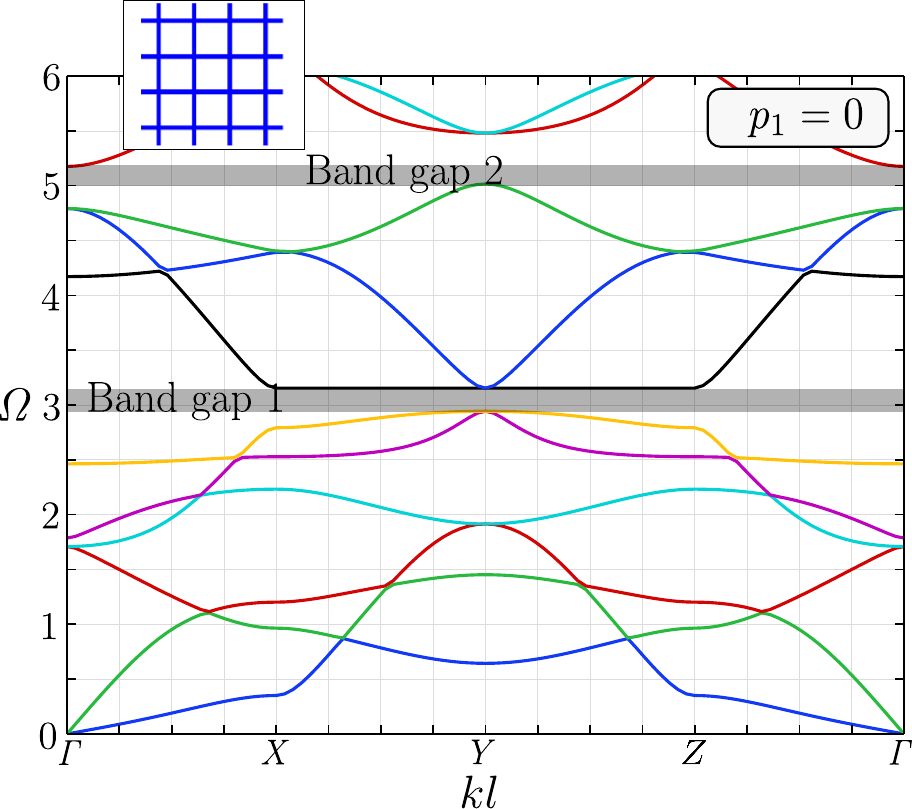}
    \end{subfigure}
    \begin{subfigure}{0.32\textwidth}        
        \includegraphics[width=0.99\linewidth]{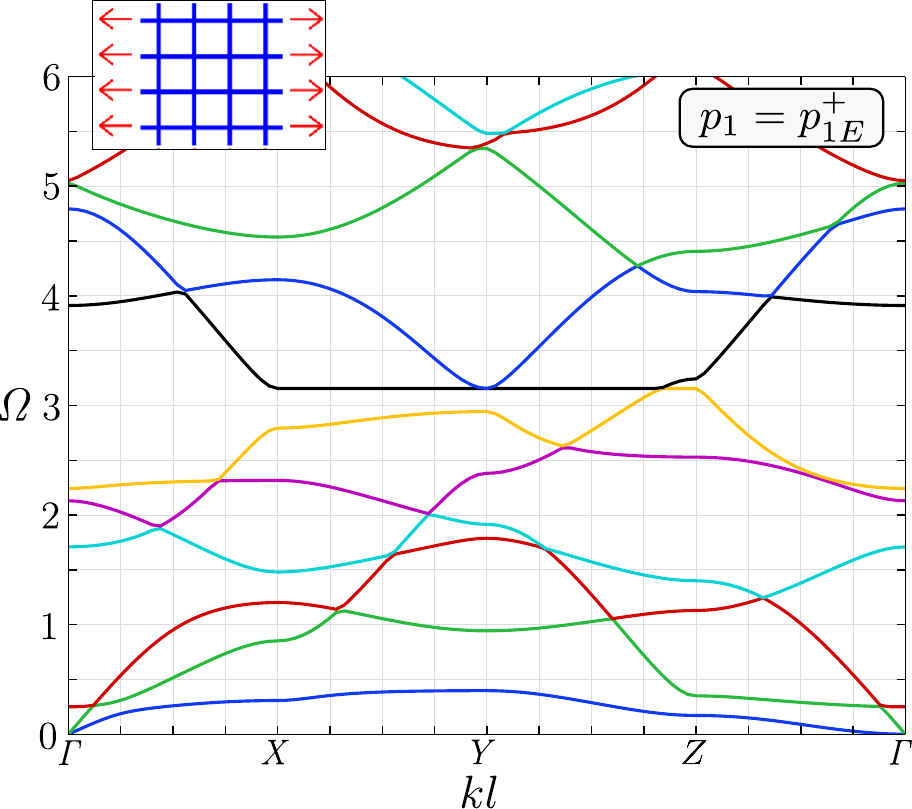}
    \end{subfigure}
    \caption{
    \label{fig:BrillonPrestres_uniaxial} Evolution of the dispersion curves with uniaxial preload  ($p_2=0$). The slider stiffness is assumed equal to $\kappa=100$. The curves on the left correspond to the compressive critical preload in Table~\ref{tab:Critical_Prestress2}, the ones in the centre correspond to the grid without preload, and the curves on the right to the grid with tensile critical preload in Table~\ref{tab:Critical_Prestress2}.
    }
\end{figure}

The dynamic response of the structure depends on both the stiffness of the sliders and the preload, the effect of the latter will be analyzed in the following to assess the stability of the structure assuming for brevity $\kappa$ = 100. Under this assumption and for three different values of preload $p_1$, including for comparison the case of null preload, the dispersion curves are shown in Fig.~\ref{fig:BrillonPrestres_uniaxial}, relative to the path $\Gamma-X-Y-Z-\Gamma$ reported in Fig.~\ref{fig:surfcut}.
%

The grid reaches a macro bifurcation in tension, on the right of the figure, and in compression, on the left, as predicted in Table~\ref{tab:Critical_Prestress2}.
When the preload achieves a critical negative or positive value (in compression or in tension), one of the acoustic surfaces exhibits a null tangent at the origin of the reciprocal space, denoting propagation of waves with infinite wavelength, that
identifies a macro instability. 
It is interesting to note that, the alteration in dispersion curve shapes is less pronounced when the lattice experiences preloading in compression compared to preloading in tension. In particular, it is possible to notice that twin band gaps persist in compression, while these disappear in tension. 

\begin{figure}[htb!]
    \begin{subfigure}{0.325\textwidth}
        \centering
        \includegraphics[width=0.95\linewidth]{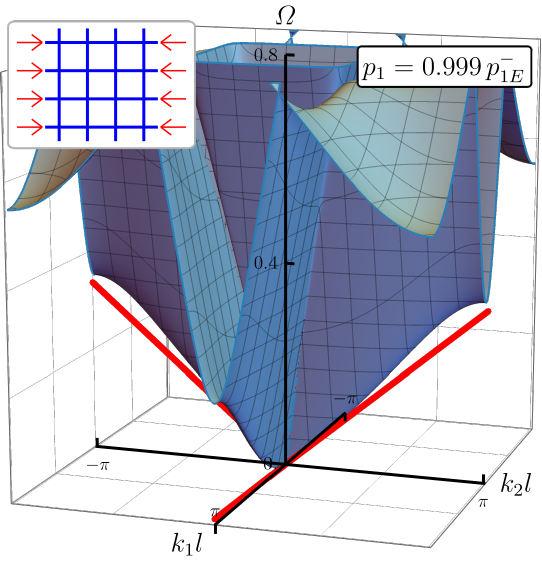}
    \end{subfigure}%
    \begin{subfigure}{0.325\textwidth}
        \centering
        \includegraphics[width=0.95\linewidth]{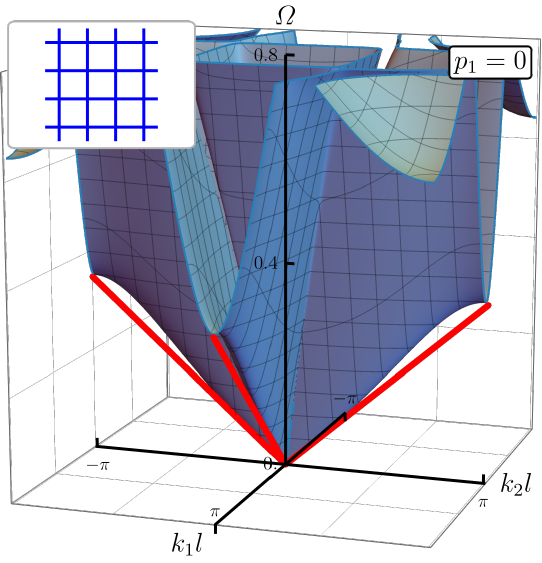}
    \end{subfigure}%
    \begin{subfigure}{0.325\textwidth}
        \centering
        \includegraphics[width=0.95\linewidth]{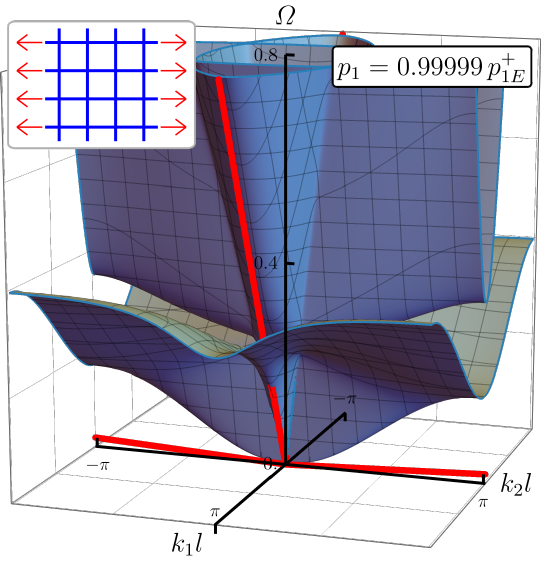}
    \end{subfigure} \\[5mm]
    \begin{subfigure}{0.325\textwidth}
        \centering
        \includegraphics[width=0.95\linewidth]{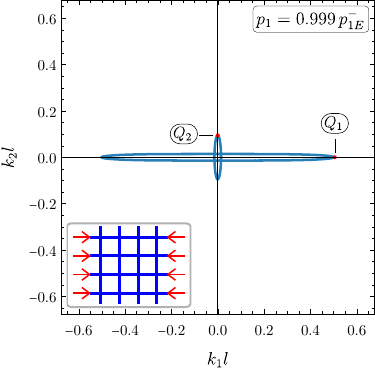}
    \end{subfigure}%
    \begin{subfigure}{0.325\textwidth}
        \centering
        \includegraphics[width=0.95\linewidth]{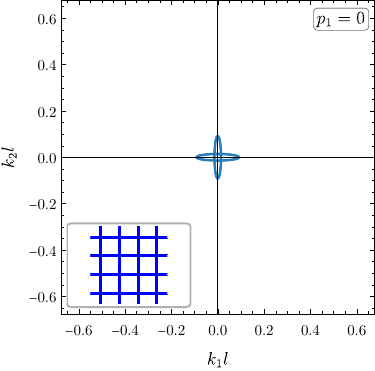}
    \end{subfigure}%
    \begin{subfigure}{0.325\textwidth}
        \centering
        \includegraphics[width=0.95\linewidth]{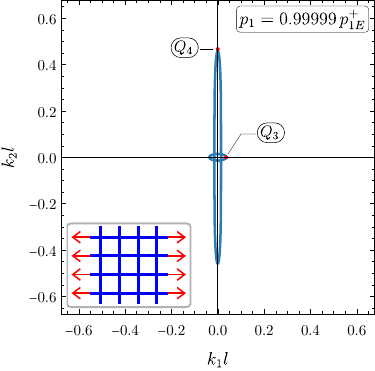}
    \end{subfigure}
    \caption{
    \label{fig:macro_bifurcation2} Dispersion surfaces (upper part) and slowness contours (lower part) at low frequency $\Omega=0.01$ for preload levels near macro bifurcation in compression (left) and tension (right), compared with the case of null preload (centre). The infinite-wavelength bifurcation occurs at the vanishing slope of one of the acoustic branches at the origin when the preload reaches the critical values reported in Table~\ref{tab:Critical_Prestress2}. Red lines are tangent to the surfaces at the origin and represent the response of the equivalent continuum.
    }
\end{figure}

The dispersion surfaces (upper part of the figure) and the relative slowness contours (lower part of the figure) at low angular frequency, $\Omega=0.01$, are reported in Fig.~\ref{fig:macro_bifurcation2} for a grid without preload (centre) and for a preload level close to the macro instability in compression (left) and in tension (right).
%

Infinite-wavelength bifurcations occur at the vanishing slope of the acoustic branches at the origin of the space $\{\Omega,\bK\}$. These tangents are reported in the figure as red lines and represent the response of the equivalent elastic continuum as obtained through homogenization, which is now recalled from \cite{bordiga_tensile_2022} and applied to the case under study.

\begin{figure}[htb!]
    \centering
    \begin{subfigure}{0.46\textwidth}
        \centering
        \caption{ $p_1=p_{1E}^-=-5.01017,\ p_2=0$}
        \includegraphics[width=0.98\linewidth]{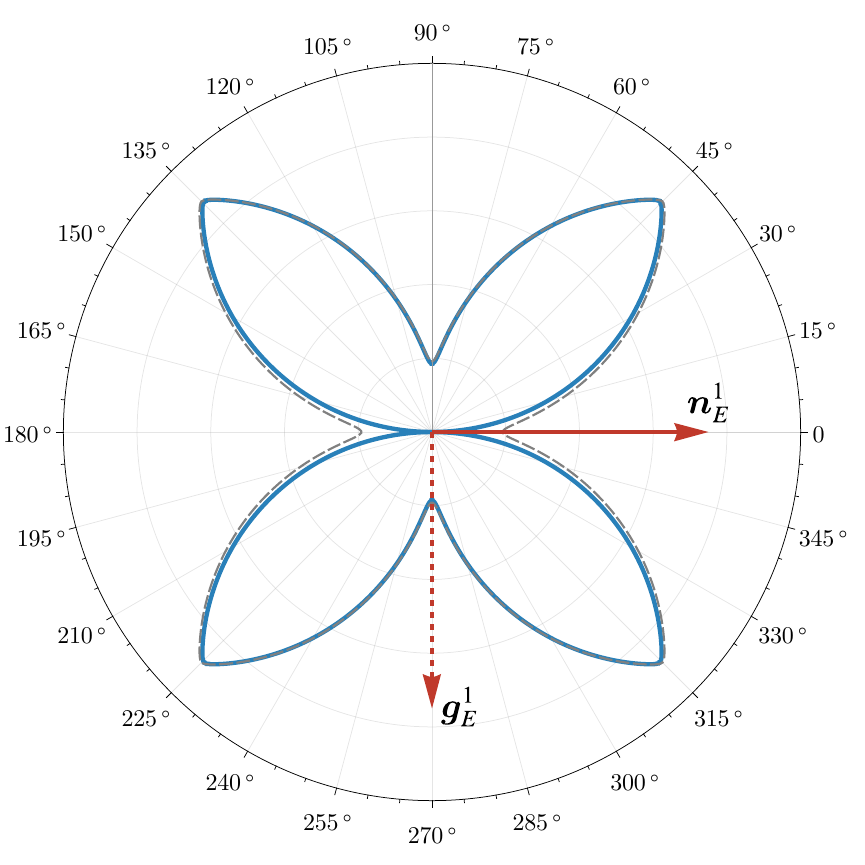}
        \label{fig:polarplot2m}
    \end{subfigure}%
    \begin{subfigure}{0.46\textwidth}
        \centering
        \caption{ $p_1=p_{1E}^+=34.066669,\ p_2=0$}
        \includegraphics[width=0.98\linewidth]{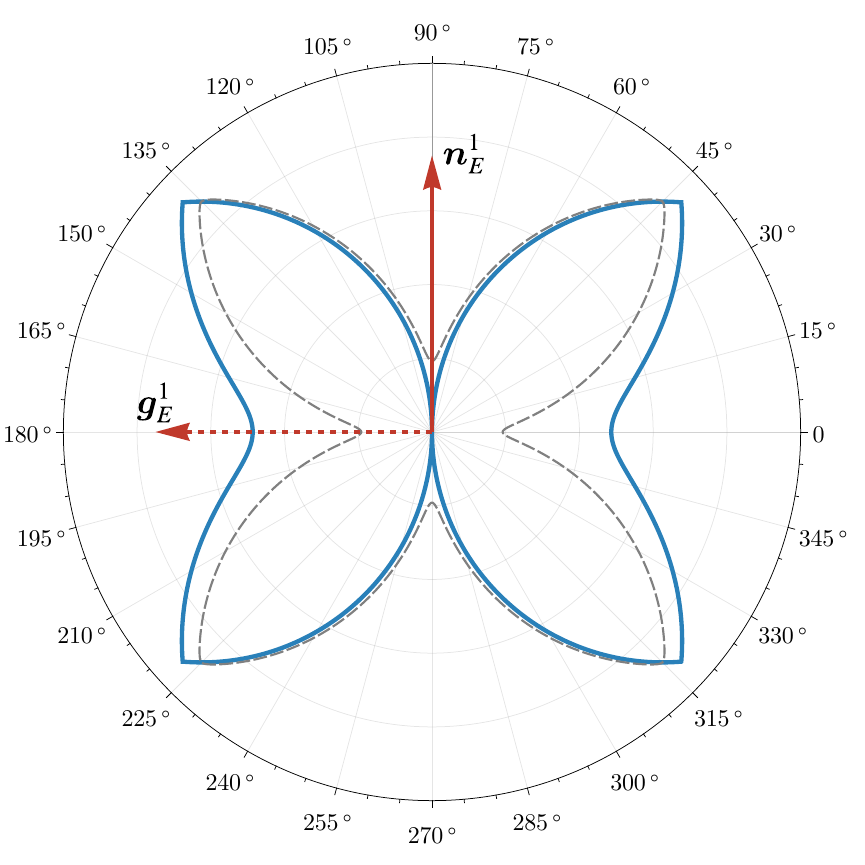}
        \label{fig:polarplot2p}
    \end{subfigure}%
    \caption{
    \label{fig:polarplot2} Polar plots of the square root of the lowest eigenvalue of the acoustic tensor of the solid equivalent to the elastic grid, as a function of the propagation direction $\bn = \cos\theta\, \be_1 + \sin\theta\, \be_2$, for a uniaxial prestress state at failure of ellipticity in compression ($p_1=p_{1E}^-=-5.01017$) and in tension ($p_1=p_{1E}^+=34.06666$), see Table \ref{tab:Critical_Prestress2}. The dashed grey line reports the behaviour of the solid without prestress, where loss of ellipticity does not occur.
    }
\end{figure}

Turning now the attention to the loss of ellipticity in the elastic material equivalent to the grid, the critical pair $\bn_E^1$ and $\bg_E^1$ can be visualized in a polar plot of the square root of the lowest eigenvalue of the acoustic tensor (of the solid equivalent to the grid) reported in Fig.~\ref{fig:polarplot2}, for two levels of uniaxial prestress, corresponding to failure or ellipticity in compression $p_{1\,E}^-$ and tension $p^+_{1\,E}$, both reported in Table~\ref{tab:Critical_Prestress2}. The plot also includes with a dashed grey line the behaviour of the equivalent continuum without prestress, where ellipticity is preserved.
%

Owing to the orthotropy inherited from the elastic grid, loss of ellipticity at the parabolic boundary in the equivalent elastic material occurs with the normal to the shear band $\bn^1_E$ parallel (orthogonal) to the preload for compression (for tension).
The associated eigenvectors $\bg^1_E$ remain orthogonal to the corresponding normals $\bn^1_E$, indicating that the modes of localization are pure shear waves, the so-called \lq shear bands'.

\begin{figure}[htb!]
    \centering
    \begin{subfigure}{0.33\textwidth}
        \centering
        \includegraphics[width=0.97\linewidth]{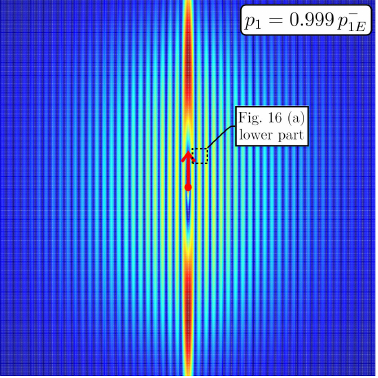}
    \end{subfigure}%
    \begin{subfigure}{0.33\textwidth}
        \centering
        \includegraphics[width=0.97\linewidth]{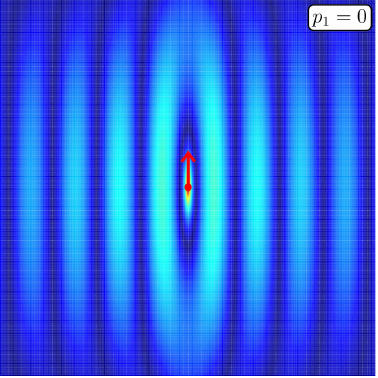}
    \end{subfigure}%
    \begin{subfigure}{0.33\textwidth}
        \centering
        \includegraphics[width=0.97\linewidth]{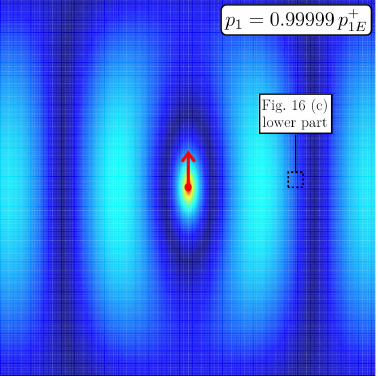}
    \end{subfigure} \\
    \begin{subfigure}{0.33\textwidth}
        \centering
        \includegraphics[width=0.97\linewidth]{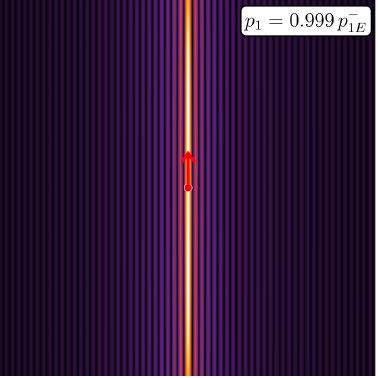}
    \end{subfigure}
    \begin{subfigure}{0.33\textwidth}
        \centering
        \includegraphics[width=0.97\linewidth]{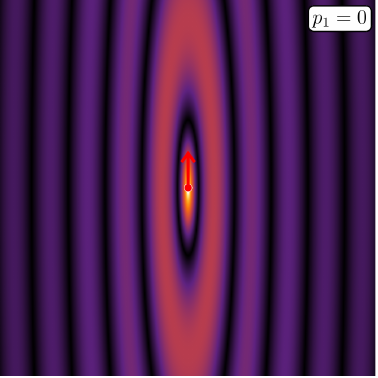}
    \end{subfigure}%
    \begin{subfigure}{0.33\textwidth}
        \centering
        \includegraphics[width=0.98\linewidth]{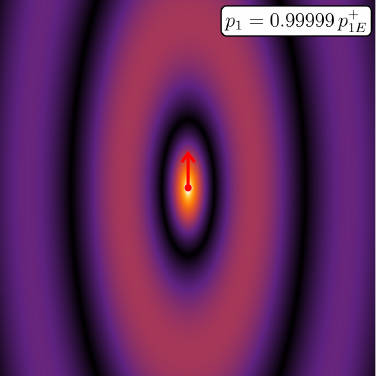}
    \end{subfigure}
    \caption{
    \label{fig:forcedver} Displacement amplitude field during vibrations of a grid of Rayleigh rods with $\kappa = 100$, excited by a vertical time-harmonic force pulsating at frequency $\Omega= 0.01$. The preload in the horizontal rods is applied up to near the critical values $p_{1\,E}^-$ and $p_{1\,E}^+$, see Table~\ref{tab:Critical_Prestress2}. The plots show shear band formation (left) near ellipticity loss, contrasted with the elliptic case (where shear bands are excluded, centre) and the other case of loss of ellipticity in tension (where shear bands do not emerge, right).
    }
\end{figure}

\begin{figure}[htb!]
    \centering
    \begin{subfigure}{0.33\textwidth}
        \centering
        \includegraphics[width=0.97\linewidth]{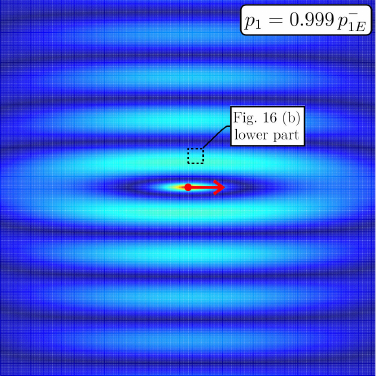}
    \end{subfigure}%
    \begin{subfigure}{0.33\textwidth}
        \centering
        \includegraphics[width=0.97\linewidth]{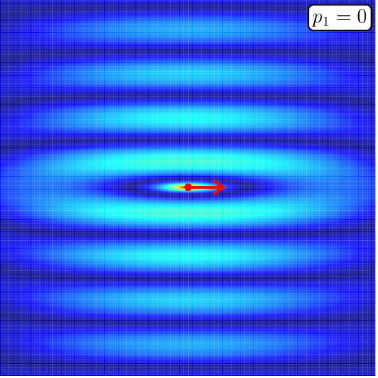}
    \end{subfigure}%
    \begin{subfigure}{0.33\textwidth}
        \centering
        \includegraphics[width=0.97\linewidth]{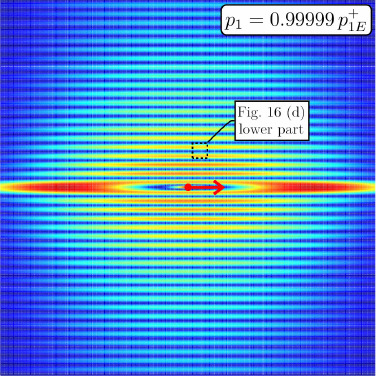}
    \end{subfigure} \\
    \begin{subfigure}{0.33\textwidth}
        \centering
        \includegraphics[width=0.97\linewidth]{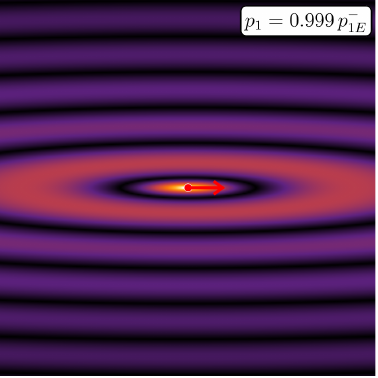}
    \end{subfigure}%
    \begin{subfigure}{0.33\textwidth}
        \centering
        \includegraphics[width=0.97\linewidth]{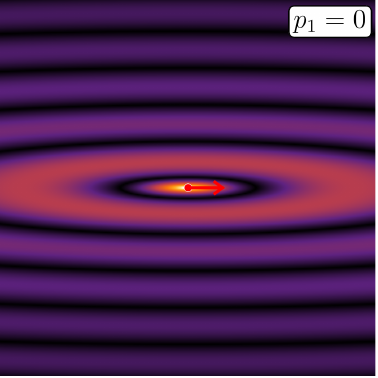}
    \end{subfigure}%
    \begin{subfigure}{0.33\textwidth}
        \centering
        \includegraphics[width=0.97\linewidth]{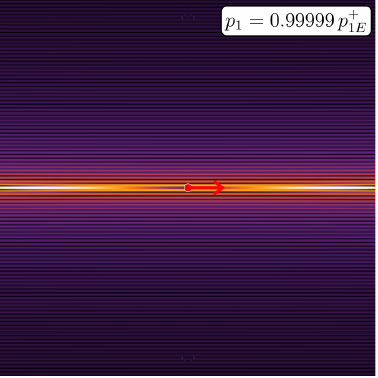}
    \end{subfigure}
    \caption{
    \label{fig:forcedhor} As for Fig.~\ref{fig:forcedver}, except that the force is horizontal. Now shear bands appear near the boundary of ellipticity loss in tension (right), but not in compression (left).
    }
\end{figure}

The dynamic response evidenced by the dispersion surfaces and slowness contours (Fig.~\ref{fig:macro_bifurcation2}) and that of the equivalent continuum are confirmed in Figs.~\ref{fig:forcedver} and \ref{fig:forcedhor}, where the behaviour of the grid (upper parts) is compared with the behaviour of the equivalent continuum (lower parts). Both the grid and the equivalent continuum are loaded through a pulsating force (applied at the central node, $C_3$ in Fig.~\ref{fig:sliding_grid}), vertical in the former figure, and horizontal in the latter. 
The behaviour of the grid is numerically obtained via f.e., while the equivalent continuum is loaded with the Green’s function obtained in \cite{piccolbigwillis}, so that an analytical solution is plotted. 
The low-frequency regime is investigated in the figures, with the concentrated forces pulsating at the dimensionless angular frequency $\Omega = 0.01$. The two preload/prestress values $0.999\,p_{E}^{-}$ and $0.99999\,p_{E}^{+}$ are analyzed, see Table~\ref{tab:Critical_Prestress2}, while the case without prestress is included for comparison. Note that to give full evidence to the shear band formation, the prestress has to be closest to the failure of ellipticity in tension (factor \lq 0.99999') than in compression (factor \lq 0.999'). 
%

Note that the two figures shown in the central columns of Fig.~\ref{fig:forcedver} are the same as in the central column of Fig.~\ref{fig:forcedhor} but rotated by $\pi/2$. 
This is not true for the other figures, due to the presence of the prestress, which is always horizontal and breaks the cubic symmetry. 

Figs.~\ref{fig:forcedver} and \ref{fig:forcedhor} show that the behaviour of the prestressed equivalent material perfectly describes the behaviour of the preloaded grid. 
The localization in a vertical (in a horizontal) shear band is evident in Fig.~\ref{fig:forcedver} (in Fig.~\ref{fig:forcedhor}) near the boundary of loss of ellipticity in compression $p_1=0.999\,p_{E}^{-}$ (in tension $p_1=0.99999\,p_{E}^{+}$). On the contrary, the particularly selected force does not excite localization, remaining unobserved in Fig.~\ref{fig:forcedver} (in Fig.~\ref{fig:forcedhor}) near the boundary of loss of ellipticity in tension $p_1=0.99999\,p_{E}^{+}$ (in compression $p_1=0.999 \,p_{E}^{-}$).
It can be concluded that for a vertical (horizontal) force, the effect of the tensile (compressive) prestress stabilizes both the grid and its equivalent continuum, so the wavelength of the displacement field increases compared to the case of the lattice without prestress.

The fact that shear bands occur parallel (orthogonal) to the direction of tensile (of compressive) prestress denotes the fact that ellipticity is lost at the parabolic boundary, \cite{bigoni_2012}. 
In this case, the localization introduces a \lq stress channelling', as noted by Pipkin \cite{pipkin}. 
Materials evidencing shear band formation at this boundary are rare. An example is the Mooney-Rivlin elasticity, but for this model, ellipticity is lost at infinite strain and is therefore unachievable. Another material showing this behaviour is the masonry-like composite introduced in \cite{muretti}, but this material is based on unilateral contacts and does not resist tension. Other possibilities are composites with stiff and parallel fibres, as preconized in \cite{pipkin}. 
Therefore, the composite here introduced opens a new way to the design of these materials.

\begin{figure}[htb!]
    \centering
    \begin{subfigure}{0.24\textwidth}
        \centering
        \caption{Point $Q_1$\label{fig:waveforms-a}}
        \includegraphics[align=c,height=0.80\linewidth]{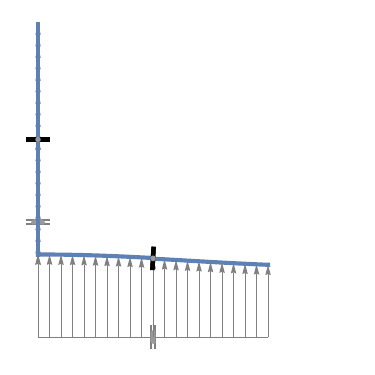}
    \end{subfigure}
    \begin{subfigure}{0.24\textwidth}
        \centering
        \caption{Point $Q_2$\label{fig:waveforms-b}}      
        \includegraphics[align=c,height=0.80\linewidth]{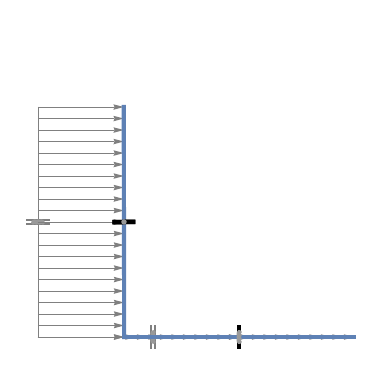}
    \end{subfigure}
    \begin{subfigure}{0.24\textwidth}
        \centering
        \caption{Point $Q_3$\label{fig:waveforms-c}}
        \includegraphics[align=c,height=0.80\linewidth]{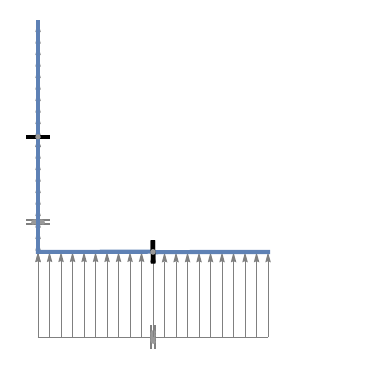}
    \end{subfigure}
    \begin{subfigure}{0.24\textwidth}
        \centering
        \caption{Point $Q_4$\label{fig:waveforms-d}}        
        \includegraphics[align=c,height=0.80\linewidth]{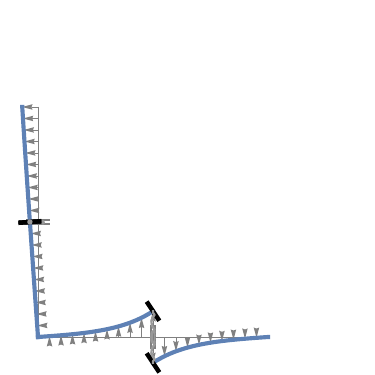}
    \end{subfigure} \\
    \centering
    \begin{subfigure}{0.24\textwidth}
        \centering        
        \includegraphics[align=c,width=0.90\linewidth]{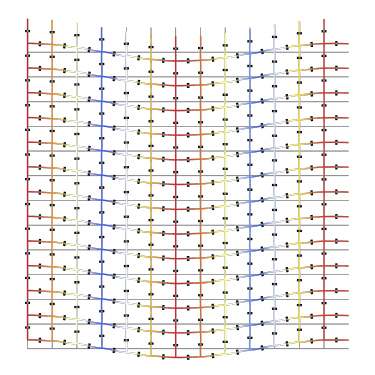}
    \end{subfigure}
    \begin{subfigure}{0.24\textwidth}
        \centering
        \includegraphics[align=c,width=0.90\linewidth]{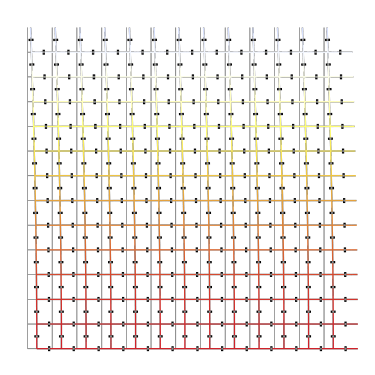}
    \end{subfigure}
    \begin{subfigure}{0.24\textwidth}
        \centering
        \includegraphics[align=c,width=0.90\linewidth]{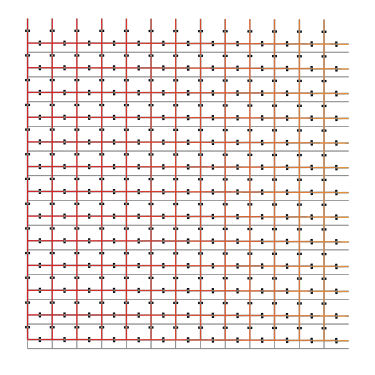}
    \end{subfigure}
    \begin{subfigure}{0.24\textwidth}
        \centering
        \includegraphics[align=c,width=0.90\linewidth]{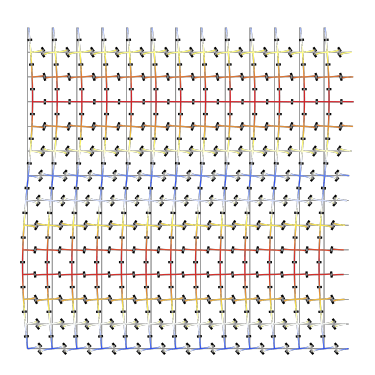}
    \end{subfigure} \\
    \begin{subfigure}{0.24\textwidth}
        \centering
        \includegraphics[align=c,width=0.80\linewidth]{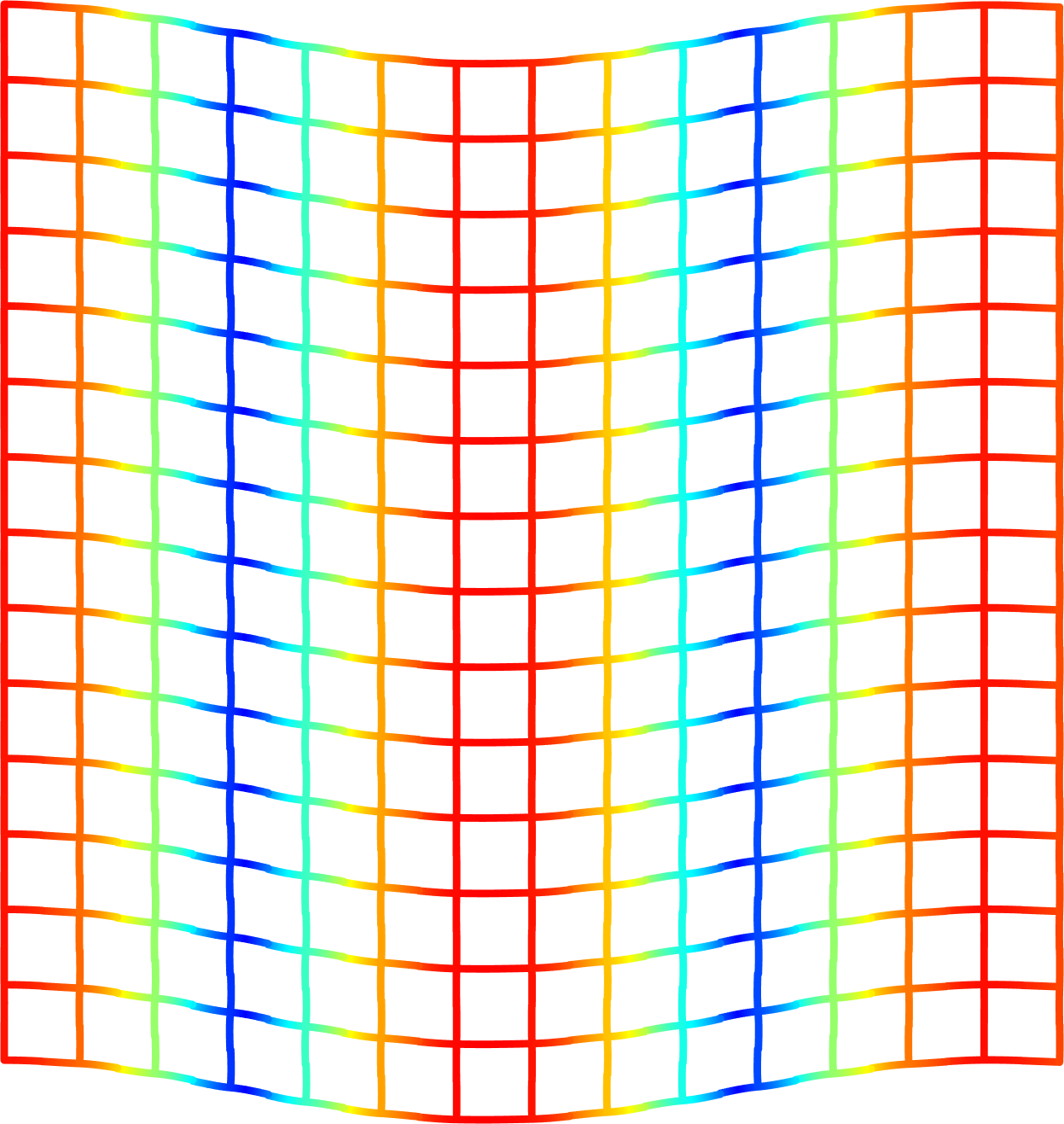}
    \end{subfigure}
    \begin{subfigure}{0.24\textwidth}
        \centering        
        \includegraphics[align=c,width=0.80\linewidth]{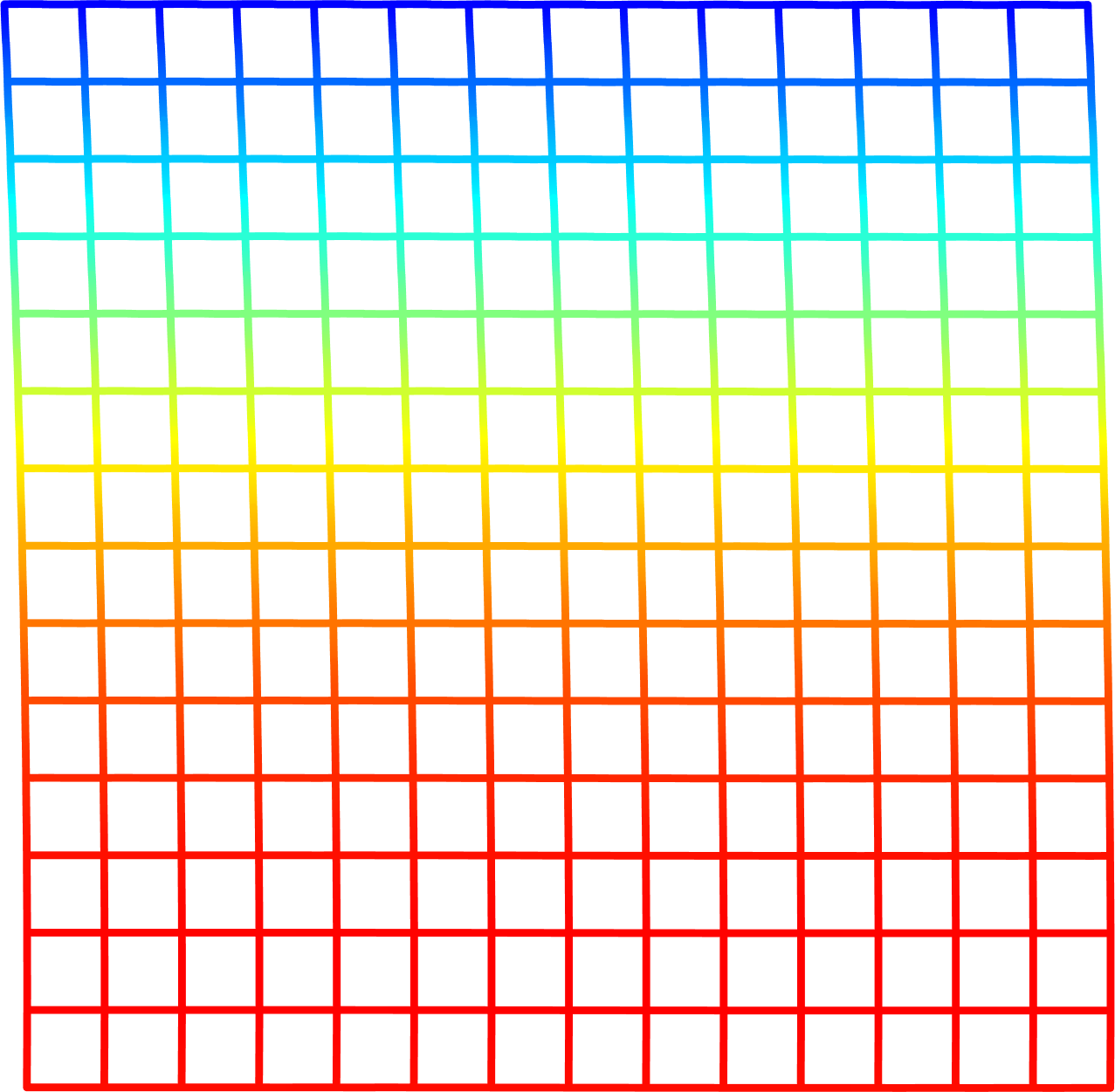}
    \end{subfigure}
    \begin{subfigure}{0.24\textwidth}
        \centering
        \includegraphics[align=c,width=0.80\linewidth]{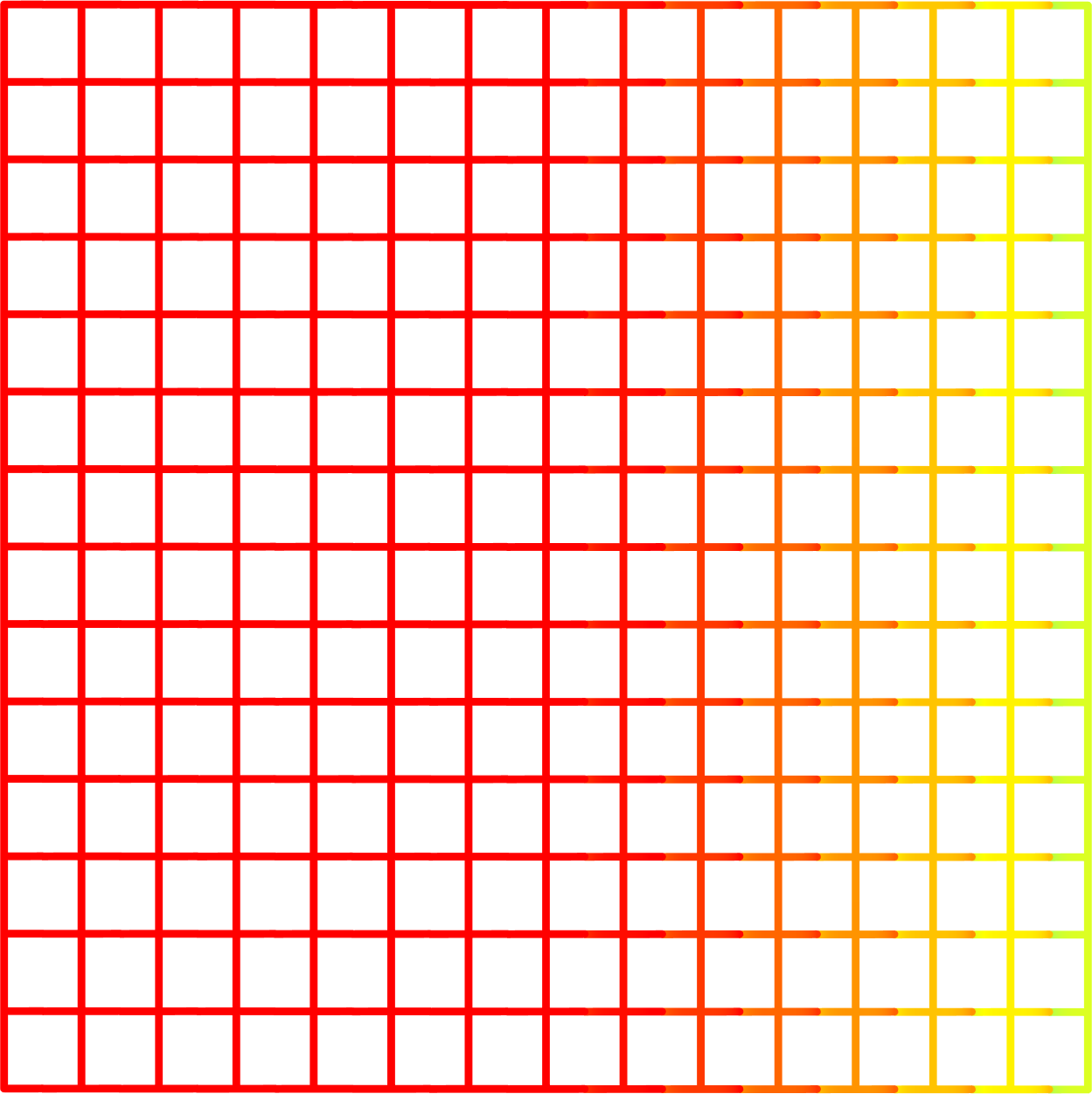}
    \end{subfigure}
    \begin{subfigure}{0.24\textwidth}
        \centering
        \includegraphics[align=c,width=0.80\linewidth]{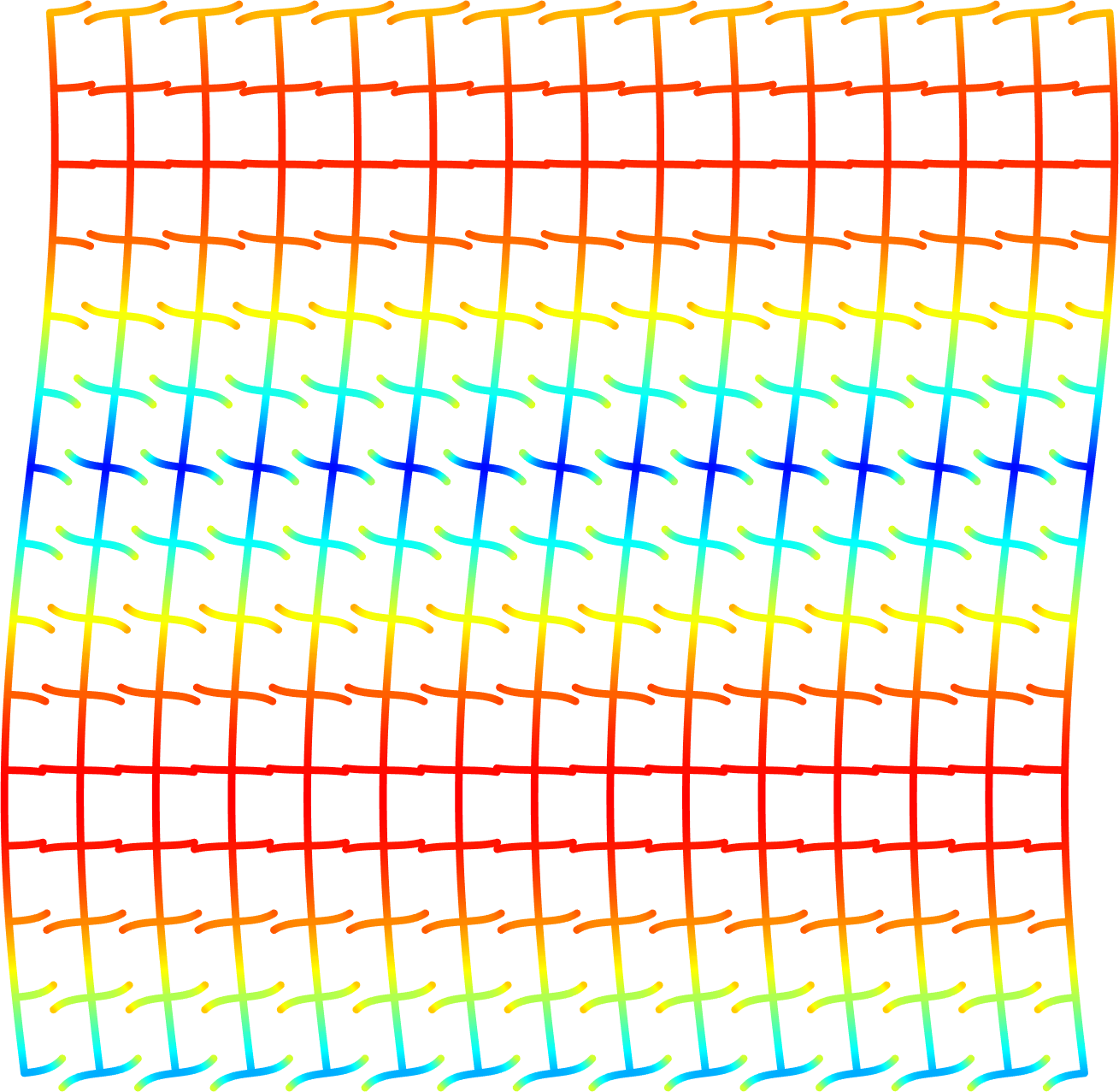}
    \end{subfigure}%
    \caption{
    \label{fig:waveforms} Selected waveforms for the grid of Rayleigh rods, for negative \subref{fig:waveforms-a}-\subref{fig:waveforms-b} and positive \subref{fig:waveforms-c}-\subref{fig:waveforms-d} values of the prestress $p_1$ near the boundary of ellipticity loss. The points $Q_i$ are shown in the slowness contours, Fig.~\ref{fig:macro_bifurcation2} lower part.
    }
\end{figure}

The behaviour of the grid and its homogenized counterpart (reported in Figs.~\ref{fig:forcedver} and \ref{fig:forcedhor}) can be better understood by considering Fig.~\ref{fig:waveforms}. 
To construct this figure, points have been selected at the ends of the branches of the slowness contours (two for critical negative prestress of compression, $Q_1$ and $Q_2$, and other two for critical positive prestress of tension, $Q_3$ and $Q_4$) shown in  Fig.~\ref{fig:macro_bifurcation2} (lower part). 
Points $Q_1$ and $Q_2$ refer to the left parts of Figs.~\ref{fig:forcedver} and \ref{fig:forcedhor}, while points $Q_3$ and $Q_4$ to the right parts, as indicated by the inset in the figures. 
For the selected points, the corresponding waveforms are reported in Fig.~\ref{fig:waveforms} for the unit cell (upper part) and for a portion of the grid (central and lower parts). 
The upper and central parts of the figure follow from the Floquet-Bloch analysis, while the lower part of the figure presents the displacement as calculated with f.e. simulations of the grid and therefore represent details of the upper parts of Figs.~\ref{fig:forcedver} and \ref{fig:forcedhor} referring to the zone marked in those figures.

The point $Q_1$ corresponds to the compressive instability: the propagation of a short-wavelength shear wave in the horizontal direction is evident; vertical rods translate longitudinally, whereas horizontal rods experience bending; all sliders are inactive. 

Points $Q_2$ and $Q_3$ correspond to a stable deformation of the lattice: the propagation of a long-wavelength shear wave, corresponding to an almost rigid translation of the rods in the horizontal and vertical directions, respectively; all sliders are inactive. 

The point $Q_4$ corresponds to the tensile instability of the lattice: the propagation of a short-wavelength shear wave in the vertical direction is visible; vertical rods undergo a rigid rotation and vertical sliders remain inactive, whereas horizontal rods are inflected and consequently horizontal sliders open. This deformation mode confirms that a tensile instability in the lattice becomes possible with the activation of the sliders, as demonstrated in \cite{zaccaria_2011}.

For the grid preloaded in compression, the propagation of the vertical shear waves, point $Q_1$, is activated when a vertical pulsating force is applied, whereas for the lattice preloaded in tension, the propagation of the horizontal shear waves, point $Q_4$, is activated by a horizontal pulsating force. In all four cases analyzed, there is a perfect match between the waveforms of Bloch waves (Fig.~\ref{fig:waveforms} central part) and the displacements of the forced lattice (Fig.~\ref{fig:waveforms} lower part). It is worth noticing, that the Fourier transform of the nodal displacements of the forced lattice, which displays the spectrum of the plane waves composing the dynamic response, nicely corresponds to the slowness contours obtained through the Floquet–Bloch analysis. For the four cases considered, the Fourier reconstruction of the signal is strongly focused around the points $Q_i$, confirming the fact that few plane waves are activated by the pulsating force. In particular, for the vertical force in compression and the horizontal force in tension, the spectra of the waves are focused around the directions of ellipticity loss, so that just the \lq slow' plane waves that are close to causing the ellipticity loss prevail in the response, as it may be expected for a material near the elliptic boundary.
%

\subsection{Equibiaxial preload $p_1=p_2$}
The lattice is subject to an equibiaxial preload state, defined by the same values of axial force in the horizontal and vertical directions $p_1=p_2$. The slider stiffnesses are again set equal to $\kappa=100$.
Occurrence of macro or micro bifurcations in the grid, corresponding to failure of ellipticity in the equivalent solid, are reported in terms of critical values of preload in Table~\ref{tab:Critical_Prestress}. 
%
\begin{table}[ht]
    \centering
    \begin{tabular}{@{}lcc@{}}
        \toprule
        \multicolumn{3}{c}{Critical values of equibiaxial preload $p_1=p_2$ for macro or micro bifurcations} \\
        \midrule
        Slider stiffness $\kappa$ & Compressive critical preload $p_E^-$ & Tensile critical  preload $p_E^+$\\
        \cmidrule(r){1-1} \cmidrule(rl){2-2} \cmidrule(l){3-3}
        1   & $-$0.687558 ~~~ (macro bif.) &  0.929147 ~~~ (micro bif.) \\
        10  & $-$2.615260 ~~~ (macro bif.) &  6.659080 ~~~ (micro bif.) \\
        100 & $-$4.741020  ~~~ (macro bif.) & 34.06669 ~~~ (micro bif.) \\
        \bottomrule
    \end{tabular}
    \caption{
    \label{tab:Critical_Prestress} Macro or micro bifurcations occurring in the elastic grid respectively in compression and tension, for different slider stiffness $\kappa$. Equibiaxial preload of the rods is assumed $p_1=p_2$. The occurrence of macro bifurcation in the grid coincides with the loss of ellipticity in the equivalent continuum, where preload becomes equivalent to prestress. Micro bifurcations remain undetected in the equivalent solid.
    }
\end{table}

The table is analogous to Table~\ref{tab:Critical_Prestress2}, except that in that table micro bifurcations are not present. The two tables share the right column, defining the same level of preload for which in one case a macro bifurcation occurs, while a micro bifurcation is critical in the other.

\begin{figure}[htb!]
    \centering
    \begin{subfigure}{0.325\textwidth}
        \centering
        \includegraphics[width=0.95\linewidth]{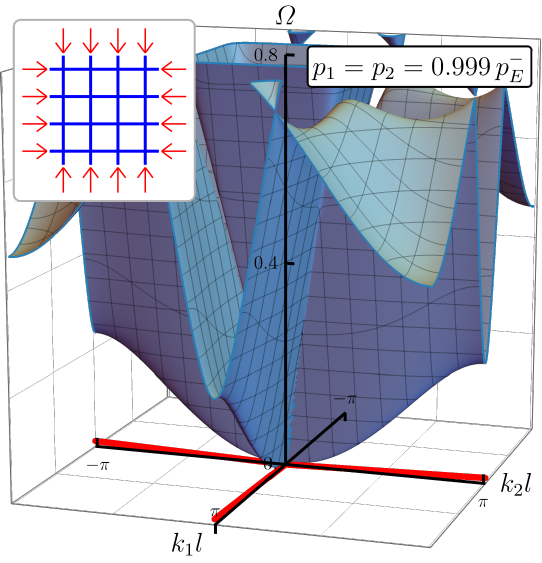}
    \end{subfigure}%
    \begin{subfigure}{0.325\textwidth}
        \centering
        \includegraphics[width=0.95\linewidth]{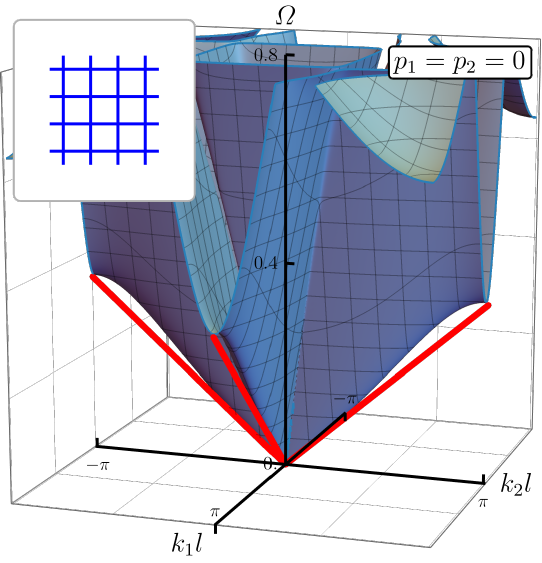}
    \end{subfigure}%
    \begin{subfigure}{0.325\textwidth}
        \centering
        \includegraphics[width=0.95\linewidth]{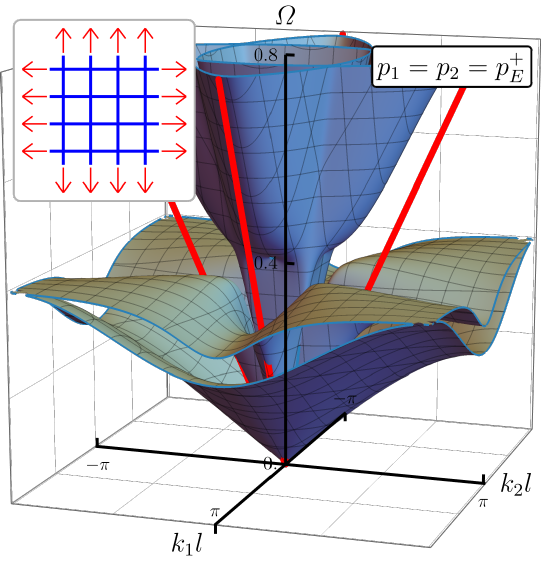}
    \end{subfigure} \\[5mm]
    \begin{subfigure}{0.325\textwidth}
        \centering
        \includegraphics[width=0.95\linewidth]{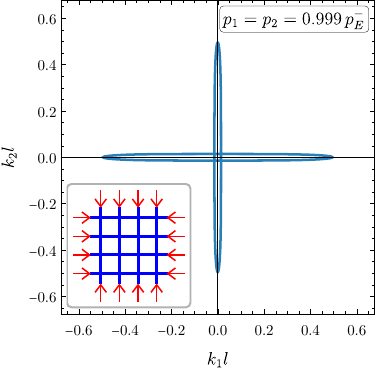}
    \end{subfigure}%
    \begin{subfigure}{0.325\textwidth}
        \centering
        \includegraphics[width=0.95\linewidth]{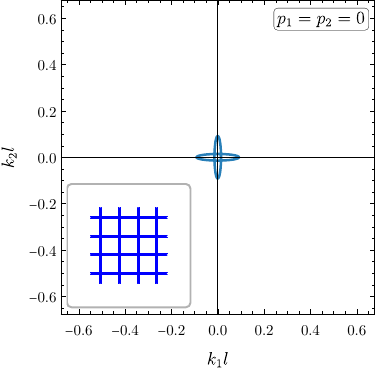}
    \end{subfigure}%
    \begin{subfigure}{0.325\textwidth}
        \centering
        \includegraphics[width=0.95\linewidth]{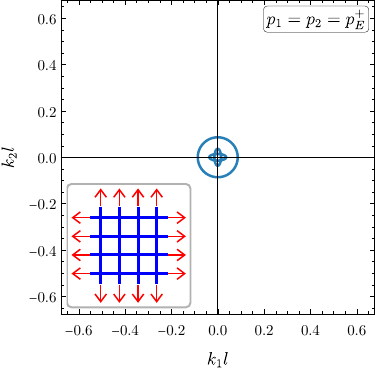}
    \end{subfigure}
    \caption{
    \label{fig:macro_bifurcation1} As for Fig.~\ref{fig:macro_bifurcation2}, except that the preload is biaxial $p_1=p_2$. Dispersion surfaces (upper part) and slowness contours (lower part) are shown at low frequency $\Omega=0.01$ for preload levels near macro bifurcation in compression (left) and micro bifurcation in tension (right), compared with the case of null preload (centre). 
    }
\end{figure}

Fig.~\ref{fig:macro_bifurcation1} is the counterpart of Fig.~\ref{fig:macro_bifurcation2} and reports the dispersion surfaces (upper part of the figure) and the relative slowness contours (lower part of the figure) at low angular frequency, $\Omega=0.01$, for a grid without preload (centre), for a preload level close to macro instability in compression (left) and micro instability in tension (right).
%

As for Fig.~\ref{fig:macro_bifurcation2}, infinite-wavelength bifurcations occur at the vanishing slope of one of the acoustic branches at the origin of the space $\{\Omega,\bK\}$ (tangents reported as red lines). 

\begin{figure}[htb!]
    \centering
    \begin{subfigure}{0.46\textwidth}
        \centering
        \caption{$p_1=p_2=p_E^-=-4.74102$}
        \includegraphics[width=0.98\linewidth]{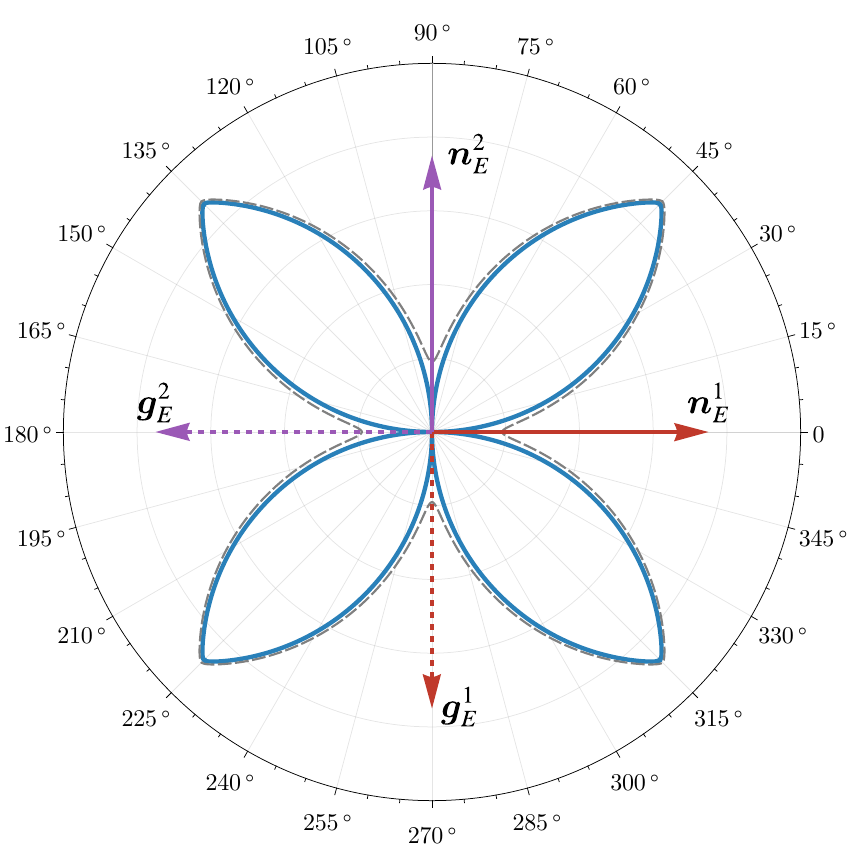}
        \label{fig:polarplot1m}
    \end{subfigure}%
    \begin{subfigure}{0.46\textwidth}
        \centering
        \caption{ $p_1=p_2=p_E^+=34.066669$}
        \includegraphics[width=0.98\linewidth]{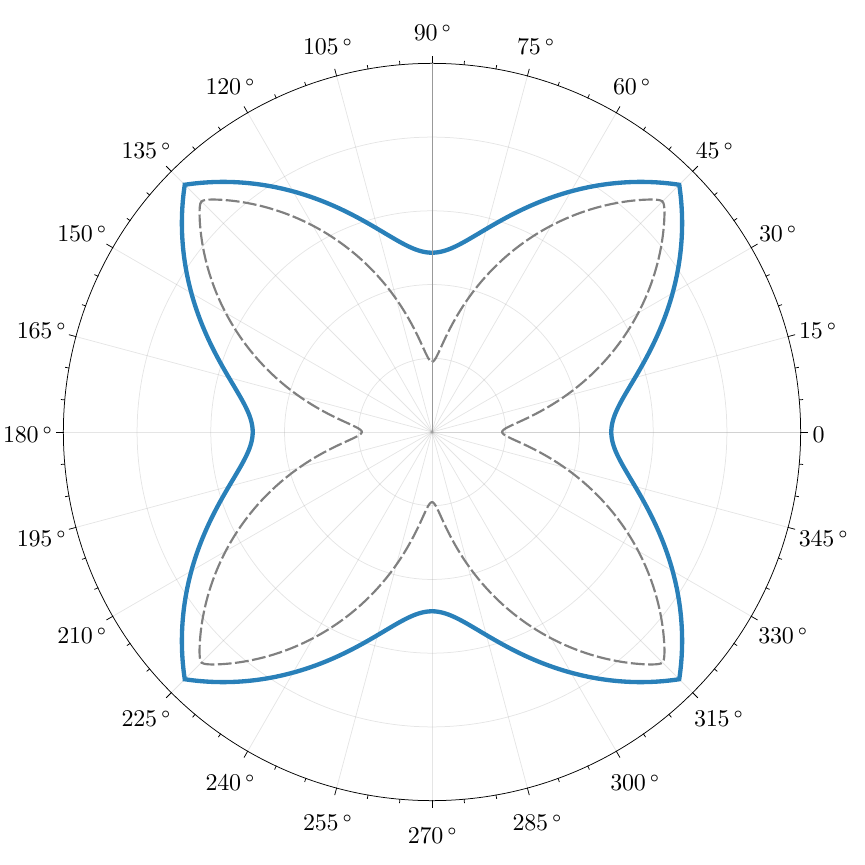}
        \label{fig:polarplot1p}
    \end{subfigure}%
    \caption{
    \label{fig:polarplot1} As for Fig.~\ref{fig:polarplot2}, except that the preload is equibiaxial, $p_1=p_2$. Polar plots are reported of the square root of the lowest eigenvalue of the acoustic tensor of the solid equivalent to the elastic grid, for an equibiaxial prestress state at failure of ellipticity in compression ($p_1=p_2=p_E^-=-4.74102$) and in tension ($p_1=p_2=p_E^+=34.06669$).
    }
\end{figure}

The critical pair $\bn_E$ and $\bg_E$, corresponding to loss of ellipticity in the equivalent elastic continuum, is reported in the polar plot of Fig.~\ref{fig:polarplot1}, analogous to Fig.~\ref{fig:polarplot2}, except that now the preload is equibiaxial, $p_1=p_2$.
%

Owing to the cubic symmetry inherited from the elastic grid, loss of ellipticity in the equivalent elastic material occurs along a pair of orthogonal directions $\bn^1_E$ and $\bn^2_E$, to which the associated wave amplitudes $\bg^1_E$ and $\bg^2_E$ are orthogonal, indicating that the modes of localization are pure shear waves, i.e.\ shear bands. 

The polar plot shown in Fig.~\ref{fig:polarplot1}-a, relates to the square lattice under the critical negative prestress $p_E^-$. Waves propagating along the horizontal and vertical directions possess equal velocities, hence ellipticity is lost when both of these velocities vanish at once, leading to two simultaneous shear bands with normals $\bn_E^1$ and $\bn_E^2$. 

When the critical positive prestress (tension) is approached, in the polar plot of Fig.~\ref{fig:polarplot1}-b the eigenvalues of the acoustic tensor do not vanish, so that ellipticity of the equivalent continuum is preserved and a micro instability becomes critical. 
This singular behaviour occurs only in the special case of equibiaxial stress. 
In this case, the slope of the acoustic branches at the origin of the space $\{\Omega,\bK\}$ does not vanish, as visible in Fig.~\ref{fig:macro_bifurcation1} on the right.
Therefore, in the case of tensile biaxial preload, a micro (instead of a macro) bifurcation of the lattice occurs, leaving the equivalent solid unaffected.

\begin{figure}[htb!]
    \centering
    \begin{subfigure}{0.33\textwidth}
        \centering
        \includegraphics[width=0.97\linewidth]{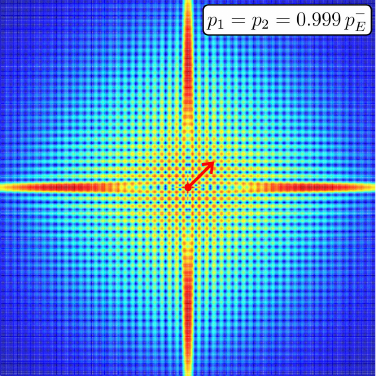}
    \end{subfigure}%
    \begin{subfigure}{0.33\textwidth}
        \centering
        \includegraphics[width=0.97\linewidth]{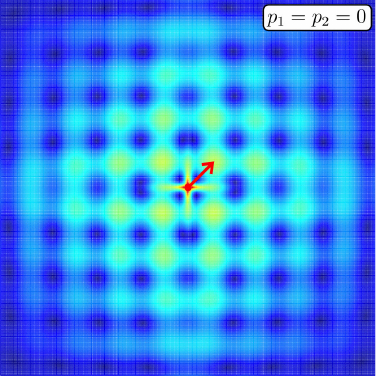}
    \end{subfigure}%
    \begin{subfigure}{0.33\textwidth}
        \centering
        \includegraphics[width=0.97\linewidth]{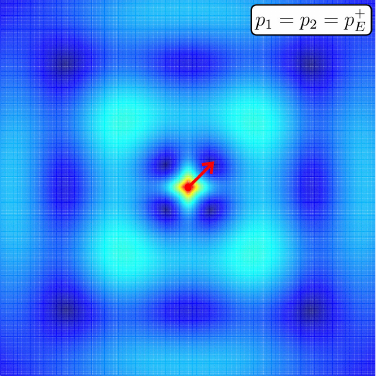}
    \end{subfigure} \\
    \begin{subfigure}{0.33\textwidth}
        \centering
        \includegraphics[width=0.97\linewidth]{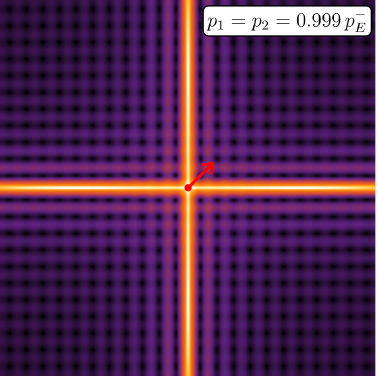}
    \end{subfigure}%
    \begin{subfigure}{0.33\textwidth}
        \centering
        \includegraphics[width=0.97\linewidth]{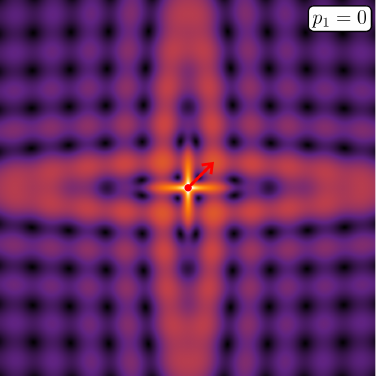}
    \end{subfigure}%
    \begin{subfigure}{0.33\textwidth}
        \centering
        \includegraphics[width=0.97\linewidth]{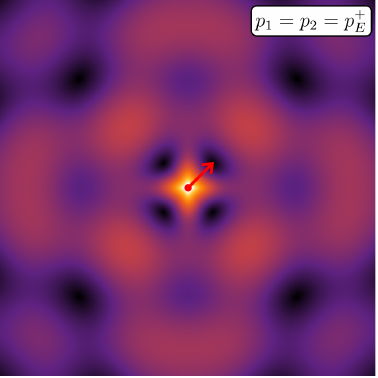}
    \end{subfigure}
    \caption{
    \label{fig:ForceP1=P2} As for Fig.~\ref{fig:forcedver}, except that the preload is equibiaxial, $p_1=p_2$ and that the concentrated force is diagonally inclined (at $\pi/4$). The displacement field is shown during vibrations of a grid of Rayleigh rods at a frequency $\Omega=0.01$, with the critical preload values, $p^-_E$ and $p^+_E$ listed in Table \ref{tab:Critical_Prestress}. Differently from the uniaxial preload, shear bands do not emerge at the failure of ellipticity in tension (on the right), where a micro bifurcation occurs simultaneously with the macro bifurcation and prevails on the latter. On the left, in compression the inclined force excites both a horizontal and a vertical shear band.}
\end{figure}

The dynamic response evidenced by the dispersion curves (Fig.~\ref{fig:macro_bifurcation1}) and that of the equivalent continuum are confirmed in Fig.~\ref{fig:ForceP1=P2}, where the behaviour of the grid (upper parts) is compared with the behaviour of the equivalent continuum (lower parts). 
Both the grid and the equivalent continuum are loaded through a pulsating force inclined at $\pi/4$ and applied at the central node ($C_3$ in Fig.~\ref{fig:sliding_grid}) in the lattice.
The low-frequency regime is investigated in the figure so that the concentrated force pulsates at the dimensionless angular frequency $\Omega=0.01$. The three preload/prestress values $0.999\,p_E^-$, $0$, and $p_E^+$ are analyzed, see Table~\ref{tab:Critical_Prestress}. 

When the prestress is compressive and near failure of ellipticity, the inclined force simultaneously excites two shear bands, orthogonal to each other, Fig.~\ref{fig:ForceP1=P2} on the left.
For a value of preload/prestress near failure of ellipticity in tension, Fig.~\ref{fig:ForceP1=P2} on the right, shear bands do not emerge. In this case, the homogenized response still captures the behaviour of the grid, but in the latter, a micro bifurcation occurs, although not activated by the point force applied at the central node. 
%

\begin{figure}[htb!]
    \centering
    \begin{subfigure}{0.33\textwidth}
        \centering
        \includegraphics[width=0.97\linewidth]{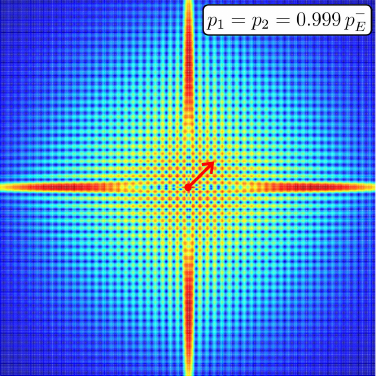}
    \end{subfigure}%
    \begin{subfigure}{0.33\textwidth}
        \centering
        \includegraphics[width=0.97\linewidth]{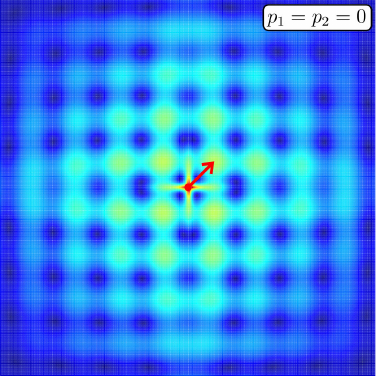}
    \end{subfigure}%
    \begin{subfigure}{0.33\textwidth}
        \centering
        \includegraphics[width=0.97\linewidth]{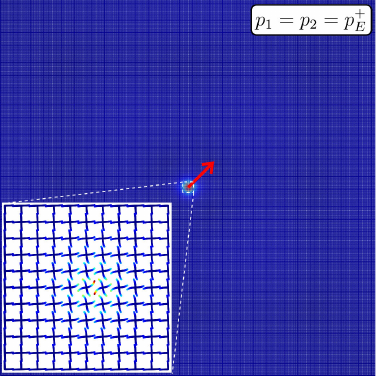}
    \end{subfigure}\\
    \begin{subfigure}{0.33\textwidth}
        \centering
        \includegraphics[width=0.97\linewidth]{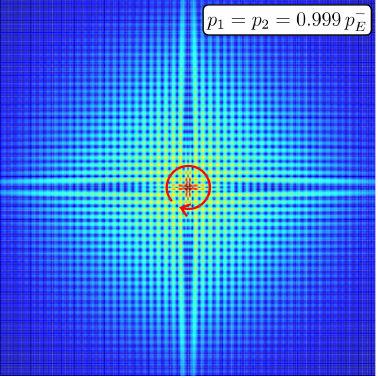}
    \end{subfigure}%
    \begin{subfigure}{0.33\textwidth}
        \centering
        \includegraphics[width=0.97\linewidth]{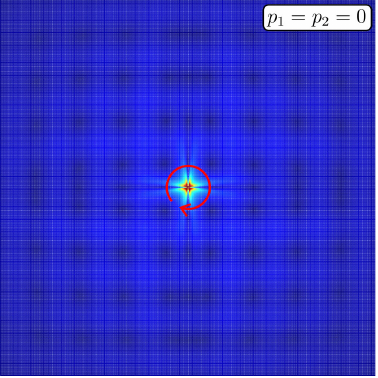}
    \end{subfigure}%
    \begin{subfigure}{0.33\textwidth}
        \centering       
        \includegraphics[width=0.97\linewidth]{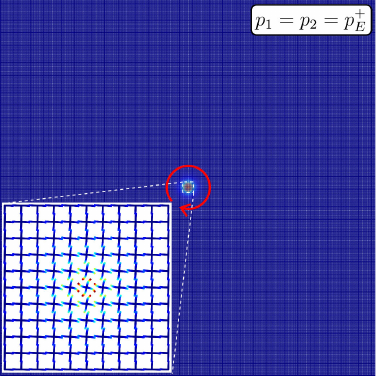}
    \end{subfigure}
    \caption{
    \label{fig:MomentP1=P2} As for Fig.~\ref{fig:ForceP1=P2}, except that the inclined force is applied on a slider face, instead of at the central node of the grid (upper part), and a concentrated moment is applied at the central node of the grid (lower part). While for compressive and null preloads the plots are very similar to those reported in Fig.~\ref{fig:ForceP1=P2} (left and centre), the results pertaining to the failure of ellipticity in tension (right) show that a micro instability prevails so that the response to the applied force or moment remains strongly localized in a vanishing small neighbourhood of the application point.
    }
\end{figure}

This situation is analyzed now in more detail focusing on the micro instability. 
The effects related to the latter can be revealed by applying the concentrated force on a face of a slider (for instance the left face of slider $S_1$ in Fig.~\ref{fig:sliding_grid}), instead of the central node of the grid (as in Fig.~\ref{fig:ForceP1=P2}), or a concentrated bending moment at the central node of the grid. 
These two cases are reported in Fig.~\ref{fig:MomentP1=P2}, only pertaining to a grid of rods numerically solved via F.E. method. Specifically, the case of a point force applied on the left face of a slider is reported in Fig.~\ref{fig:MomentP1=P2} upper part, whereas the case of a moment applied at the central node is reported in Fig.~\ref{fig:MomentP1=P2} lower part.

Fig.~\ref{fig:MomentP1=P2} reveals that for compressive and null preload the situation remains very similar to that shown in Fig.~\ref{fig:ForceP1=P2} (on the left and the central part). However, there is a striking difference when the microinstability occurs (compare Fig.~\ref{fig:ForceP1=P2} on the right with Fig.~\ref{fig:MomentP1=P2} on the right). In the latter case, the loss of compliance induced by micro instability strongly limits the effects of the concentrated force and moment, which tend to vanish, namely, the bending and the node rotations extend to a limited number of rods in the close neighbourhood of the application point. 
%

\FloatBarrier
\newpage

\section{Concluding remarks}
\label{sec:concluding}
A rigorous Floquet-Bloch analysis and homogenization of periodic grids of preloaded elastic rods have shown that the inclusion of slider constraints introduces strong mechanical effects. These are related to the dynamics of the grid and the occurrence of localization of deformation in the form of shear bands. These localizations occur both in compression and in tension, so that materials working like thin sheets can be obtained and subjected to tensile loads. 
Metamaterials with embedded sliders can be designed in a way that shear bands emerge at the elliptic/parabolic boundary, evidencing a single shear band. In this way, stress channelling and vibration localization can be realized. These effects can be used to propagate signals in preferred directions within a plane, with minimal dispersion. The localization was shown to be
activated or deactivated simply increasing or decreasing the axial preload of the rods. Therefore, the presented results open new possibilities in the mechanics of metamaterials.

\appendix

\section{Full expression for the effective acoustic tensor}
\label{sec:effective_acoustic_tensor}
The complete analytic expression for the effective acoustic tensor of the lattice analyzed in Section~\ref{sec:formulation_problem} is here reported in the dimensionless form $\bA = \frac{EA_1}{l_1}\, \Bar{\bA}$.
\begin{align*}
    \Bar{A}_{11} &= \frac{k_1^2}{\xi} + \frac{a_{11}k_2^2}{\lambda_2^2D} \,, \\
    \Bar{A}_{22} &= \chi k_2^2 + \frac{a_{22}k_1^2}{\lambda_1^2 \xi D} \,, \\
    \Bar{A}_{12} &= \Bar{A}_{21} = k_1 k_2 \frac{a_{12}}{D} \,,
\end{align*}
where
\begin{multline*}
    a_{11} = \\
    \lambda_1^2 \xi \chi^2 p_2^2 
    \left[e^{\sqrt{p_1}} \left(\sqrt{p_1}-2\right) \left(p_1^{3/2}-2\kappa_1\right) - \left(\sqrt{p_1}+2\right) \left(p_1^{3/2}+2\kappa_1\right)\right] 
    \left[e^{\sqrt{p_2}} \left(p_2^{3/2}-2\kappa_2\right) + p_2^{3/2}+2\kappa_2\right] \\
    + \lambda_2^2 \chi p_1 p_2^{3/2} 
    \left[e^{\sqrt{p_2}} \left(p_2^{3/2}-2\kappa_2\right) - \left(p_2^{3/2}+2\kappa_2\right)\right] 
    \left[e^{\sqrt{p_1}} \left(p_1^{3/2}-2\kappa_1\right) + p_1^{3/2}+2\kappa_1\right] \,,
\end{multline*}
\begin{multline*}
    a_{22} =
    \lambda_1^2 \xi \chi p_1^{3/2} p_2 
    \left[e^{\sqrt{p_1}} \left(p_1^{3/2}-2\kappa_1\right) - \left(p_1^{3/2}+2\kappa_1\right)\right] 
    \left[e^{\sqrt{p_2}} \left(p_2^{3/2}-2\kappa_2\right) + p_2^{3/2}+2\kappa_2\right] \\
    + \lambda_2^2 p_1^2 
    \left[e^{\sqrt{p_2}} \left(\sqrt{p_2}-2\right) \left(p_2^{3/2}-2\kappa_2\right) - \left(\sqrt{p_2}+2\right) \left(p_2^{3/2}+2\kappa_2\right)\right] 
    \left[e^{\sqrt{p_1}} \left(p_1^{3/2}-2\kappa_1\right) + \left(p_1^{3/2}+2\kappa_1\right)\right] \,,
\end{multline*}
\vspace{3mm}
\begin{equation*}
    a_{12} = 
    2 \chi p_1 p_2 
    \left[e^{\sqrt{p_1}} \left(p_1^{3/2}-2\kappa_1\right) + p_1^{3/2}+2\kappa_1\right] 
    \left[e^{\sqrt{p_2}} \left(p_2^{3/2}-2\kappa_2\right) + p_2^{3/2}+2\kappa_2\right] \,,
\end{equation*}
\begin{multline*}
    D = \\
    \lambda_1^2 \xi \chi p_2 
    \left[e^{\sqrt{p_1}} \left(\sqrt{p_1}-2\right) \left(p_1^{3/2}-2\kappa_1\right) - \left(\sqrt{p_1}+2\right) \left(p_1^{3/2}+2\kappa_1\right)\right] 
    \left[e^{\sqrt{p_2}} \left(p_2^{3/2}-2\kappa_2\right) + p_2^{3/2}+2\kappa_2\right] \\
    + \lambda_2^2 p_1 
    \left[e^{\sqrt{p_2}} \left(\sqrt{p_2}-2\right) \left(p_2^{3/2}-2\kappa_2\right) - \left(\sqrt{p_2}+2\right) \left(p_2^{3/2}+2\kappa_2\right)\right] 
    \left[e^{\sqrt{p_1}} \left(p_1^{3/2}-2\kappa_1\right) + p_1^{3/2}+2\kappa_1\right] \,.
\end{multline*}

\section*{Acknowledgements}
All the authors acknowledge financial support from the European Research Council (ERC) under the European Union’s Horizon 2020 research and innovation programme (Grant agreement No. ERC-ADG-2021-101052956-BEYOND).

\printbibliography

\end{document}